\input amstex
\input xy
\xyoption{all}
\documentstyle{amsppt}
\document
\magnification=1200
\NoBlackBoxes
\nologo
\hoffset1.5cm
\voffset2cm
\vsize15.5cm
\def\F{\bold{F}}

\def\C{\bold{C}}

\def\Q{\bold{Q}}

\def\Z{\bold{Z}}

\def\F{\bold{F}}

\def\cH{\Cal{H}}
\def\cC{\Cal{C}}
\def\cO{\Cal{O}}
\def\cV{\Cal{V}}
\def\cS{\Cal{S}}
\def\cG{\Cal{G}}

\def\cL{\Cal{L}}
\def\cT{\Cal{T}} 
 
\def\cD{\Cal{D}} 
\def\cE{\Cal{E}} 
\def\cB{\Cal{B}} 
\def\cI{\Cal{I}} 

\def\Tr{\roman{Tr}}
\def\GL{\roman{GL}}
\def\Heis{\roman{Heis}}

\def\fm{\bold{m}}

\bigskip


\bigskip

\centerline{\bf MOUFANG PATTERNS}

\medskip

\centerline{\bf AND  GEOMETRY OF INFORMATION}

\medskip

\centerline{\bf No\'emie Combe,\quad  Yuri I.~Manin, \quad  Matilde Marcolli}

\bigskip

\centerline{\it Dedicated to Don Zagier}

\bigskip

{\it ABSTRACT.}  Technology of data collection and information transmission 
is based on various mathematical models of encoding. The words ``Geometry of information''
refer to such models, whereas the words ``Moufang patterns''
refer to various sophisticated symmetries
appearing naturally in such models.

\smallskip

In this paper we show that the symmetries of spaces of probability distributions,
endowed with their canonical Riemannian metric of information geometry, 
have the structure of a commutative Moufang loop. We also show that the
$F$--manifold structure on the space of probability distribution can be
described in terms of differential $3$--webs and Malcev algebras.  We then
present a new construction of (noncommutative) Moufang loops associated to
almost--symplectic structures over finite fields, and use then to construct
a new class of code loops with associated quantum error--correcting codes
and networks of perfect tensors.

\medskip



\bigskip

\centerline{\bf 0. Introduction and Summary}

\medskip

This paper can be roughly subdivided into two parts: Sections 1--4
and Sections 5--6. 

\smallskip

The words "Geometry of information''  in the first part refer 
to models of databases subject to noise -- {\it probability
distributions on finite sets.}  The same words in the
second part refer to theory of {\it error--correcting codes.} 

\smallskip

The introductory subsections of each part define mathematical
structures, describing symmetries of relevant geometries:
Commutative Moufang Loops in the first part,
and (virtually) noncommutative Moufang Loops in the second part.

\smallskip

Here are some more details.

\smallskip

In Sec.~1 we recall the definitions of
symmetric quasigroups and CH--quasigroups, describe their relations to
commutative Moufang loops, and we summarise their role in algebraic
geometry, as in [Ma86], in particular in the case of the set of algebraic points of
a cubic curve in the projective plane. In Sec.~2 we give 
the differentiable version of Moufang loops, in the form of 
Malcev algebras [Mal55], which generalise to loops the relation between 
Lie algebras and Lie groups. 

\smallskip

The main new results of this paper
start in Sec.~3, where we consider spaces of
probability distributions on  finite sets, endowed with a
family of canonical Riemannian metrics. We consider symmetries
of the space of probabilities given by automorphisms of order
two that are boundary limits of the reflections of geodesics
about the center. We show that these automorphisms define a
composition law on the set of points that is an abelian
symmetric quasigroup. 

\smallskip

In Sec.~4 we consider the structure of $F$--manifold on the
space of probability distribution, previously discussed in [CoMa20]
and [CoMaMar21]. Using a formulation in terms of differential 
$3$--webs, we show the compatibility between the quasigroup
structure, which determines a family of (bundles) of Malcev algebras, 
and the family of $F$--structures, whose flat structures are obtained
from the Chern connections of the family of differential $3$--webs. 

\smallskip

In Sec.~5 we recall some definitions and properties
of (noncommutative) Moufang loops, from [Gri86] and [Hsu00a].
We also recall some notions and results about
classical error--correcting codes, from [Ma12], [MaMar12].
Then we introduce code loops, as defined in [Gri86] (see also [Conw85]) and 
further studied in [Hsu00a].

\smallskip

In Sec.~6 we develop a construction of code loops
based on a generalisation to the almost--symplectic case
of the symplectic quantisation in positive characteristic
of [GuHa09] and [GuHa12]. In particular, we extend to
the case of code loops results of [HMPS18] on the
construction of quantum error--correcting codes and
perfect tensors from isotropic and Lagrangian subspaces,
in the symplectic quantisation case. 

\smallskip

More precisely, at the beginning
of Sec.~6 we recall in more details the construction of 
code loops of [Gri86] and [Hsu00a] and its formulation
in terms of doubly even binary codes. In Sec.~6.1 we
recall some definitions and results on quantum error-correcting codes.
In Sec.~6.2 we recall the result of [HMPS18] on the
symplectic CRSS algorithm constructing quantum codes
from classical codes that are isotropic subspaces of
a symplectic vector space, via the quantisation procedure
of [GuHa09] and [Gu Ha12].

\smallskip

Similarly, in Sec.~6.3 we
recall the result of of [HMPS18] on the construction of
perfect tensors from Lagrangian subspaces in general
position with respect to the Darboux decomposition. 
In Sec.~6.4 we show how the results summarised in
Sec.~6.2 and 6.3 extend to the case of characteristic $2$
that was not treated in [HMPS18]. 

\smallskip

In Sec.~6.5 we
introduce a new construction of code loops based on
almost-symplectic vector spaces over finite fields.
We describe the Moufang condition for these loops 
in terms of the construction of [GriPir18] of free 
Moufang loops in the variety generated by code loops.
In Section~6.6
we prove that the symplectic CRSS algorithm for
constructing quantum codes from classical codes
extends to the case of almost--symplectic code loops. 
In Sec.~6.7 we show that the construction of perfect 
tensors from Lagrangians also extends to the case of 
almost--symplectic code loops, when 
the almost-symplectic structure is locally conformally
symplectic. This condition ensures that a version of
the Darboux decomposition can still be obtained,
hence the general position property of Lagrangians
that gives the perfect tensor condition. In Sec.~6.8
we discuss networks of perfect tensors and the
associated entanglement entropy function. We also
recall the equivalence of categories between loops
and Latin square designs and the subcategory of
central Latin square designs that corresponds to Moufang loops.

\smallskip

Finally, in Sec.~6.9 we show
that the construction of perfect tensors obtained in
Sec.~6.7 determines a tensor network on a subgraph
of the graph associated to the Latin square design
of the almost-symplectic code loop. In Sec.~6.10
we recall the formalisms of chamber systems associated to loops
and their Latin square designs and their relation
to buildings, and formulate some questions on the
possible construction of tensor networks on these
chamber systems  and buildings and their possible
holographic properties.

\bigskip

\centerline{\bf 1. Quasigroups, commutative Moufang loops,} 

\smallskip
\centerline{\bf and algebraic varieties}

\medskip

{\bf 1.1. Symmetric quasigroups.} As in [Ma86], Ch. 1, we start with considering a
set $E$ with binary composition law $\circ : E\times E \to E, (x,y)\mapsto x\circ y.$
Such a structure will be called {\it a symmetric quasigroup} if the triple relation
$L(x,y,z): x\circ y=z$ is $\bold{S}_3$--invariant.

\smallskip

A symmetric quasigroup $(E,\circ )$  as above is called {\it abelian},
if for any element $u\in E$ the composition law $ (x,y) \mapsto u\circ (x\circ y)$
turns $E$ into an abelian group with identity $u$.

\smallskip

Finally, a symmetric quasigroup $(E, \circ )$ is called a {\it CH--quasigroup},
if any subset of $E$ of cardinality $3$ generates an abelian subquasigroup.
(An explanation of $CH$ in this definition will be given below).

\medskip

{\bf 1.2. Commutative Moufang loops (CML).} By definition, a CML is a set $E$ endowed with  a commutative
binary composition law $*: E\times E \to E: (x,y)\mapsto x*y$, with identity
$u\in E$ and left inverse map $E\to E: x\mapsto x^{-1}$. The main additional constraints
below were called ``weak associativity'' relations in Def. 1.4 of [Ma86]:
$$
x*(x*y) = (x*x)*y, (x*y)*(x*z) = (x*x)*(y*z),
$$
$$
(x*y)*(x*z)= ((x*x)*y)*z.
$$
The {\it associative centre} of a CML $(E,*)$ is the subset
$$
Z(E):= \{x\in E\,|\, x*(y*z) = (x*y)*z \ for\  all\  y, x\in E\}.
$$
Together with induced multiplication, $Z(E)$ is an associative subloop, and therefore
an abelian group. The quotient loop $E/Z(E)$ is a CML of exponent 3:
$$
x^{*3} := x*(x*x) = (x*x)*x = u.
$$

\smallskip
Loops of exponent 3 form a subcategory of  all CML's. 

\medskip

{\bf 1.3. Connections between quasigroups and Moufang loops.} There are several
natural ways to make CH--quasigroups, resp. CML's, objects of categories, by defining
morphisms between them. Then connections between objects of these two categories must naturally become functors.
But in this subsection, we will  neglect morphisms.

\medskip

{\bf 1.3.1. Proposition.} {\it (i) Let $(E,\circ)$ be a CH--quasigroup, and $u\in E$ its element.
Then $E$ endowed with composition law $(x,y)\mapsto x*y:= u\circ(x\circ y)$ is a CML with identity $u$.
Different choices of $u$ lead to isomorphic CML's.}

\smallskip

{\it (ii) Let $(E,*)$ be a CML, and $c$ an element of its associative centre $Z(E)$.
Then $E$ with composition law $x\circ y:= c*x^{-1}*y^{-1}$ is a CH--quasigroup.}

\medskip

{\bf 1.4. Quasigroups and loops in algebraic geometry.} Appearance of the simplest CML's
in algebraic--geometric setup was motivated in [Ma86] by smooth cubic curves in a projective
plane $\bold{P}^2_K$  over a field $K$. The set $E$ of $K$--points of such a curve $X$  forms a CML
with composition law $x*y=u\circ (x\circ y)$ as in Prop. 1.3.1 (i) above, if $u+x+y$
is the intersection cycle of $X$ with a projective line $\bold{P}^1_K \subset \bold{P}^2_K$.

\smallskip

Further developments led to a theory of such structures,
necessary for their applicability to higher--dimensional cubic hypersurfaces. 

\smallskip

Below we will argue
that finitely generated loops of this type naturally act as symmetries of
spaces of probability distributions of finite sets. Therefore, it is worth considering
posets in groupoids and the related thin categories of such loops,
morphisms in which are embeddings: see [CoMa21], subsections 5.1 and 5.2.

\bigskip

\centerline{\bf 2. Analytical commutative Moufang loops and  Malcev algebras.}  

\medskip

In his work [Mal55],
I.  A. Malcev considered Moufang loops endowed with an additional structure
of (local) differentiable or (real) analytic variery, with which the Moufang
composition is compatible. He has introduced and studied the induced
structures upon tangent vector bundles, generalising
the relationship between Lie groups and Lie algebras.

\smallskip

Below we will sketch this theory. Our exposition relies mainly
upon [Sa61], [Na88], [Na92], and [Pa03].

\smallskip

{\bf 2.1. Definition} (Def. 4.2 in [Na92]). {\it A vector space $\Cal{T}$ over 
a field, endowed with an antisymmetric bilinear composition map
$$
[,]^\mu: \, L\otimes L \to L, \quad x\otimes y \mapsto [x,y]^\mu
$$
is called a Malcev algebra, it satisfies the identity
$$
[[x,y]^\mu, [x,z]^\mu]^\mu= [[[x,y]^\mu, z]^\mu,x]^\mu + [[[y,z]^\mu,x]^\mu,x]^\mu
 + [[[z,x]^\mu, x]^\mu, y]^\mu.   \eqno(2.1)
$$
}

{\bf 2.1.1. Remarks.} (i) We replaced the notation $[,]$ of [Na92] by our $[,]^{\mu}$  in order
to distinguish it from the usual Lie brackets for vector fields.
Moreover, in our applications, the ground field
will be mostly real or complex numbers.
\smallskip

(ii) The same definition is given in Sec. 2 of [Sa61], but our $[x,y]^\mu$
is denoted there $xy$, or $x.y$, or $(x.y)$ (cf. (2.1) and (2.2)).

From the operadic perspective, Nagy's notation is more consistent.

\medskip

More general (not necessarily commutative) Moufang loops will also be discussed
in the following sections of this paper. These are sets $\cL$ endowed with
a binary composition law with identity and inverse element, 
that satisfies a weak associativity relation (Moufang identity): see 
Definition~5.1.1 below. 

\medskip

{\bf 2.2. Proposition.} {\it Let $(\cL,*)$ be a Moufang loop with identity $e$, endowed
with compatible structure of real analytic (or smooth) variety.

Then the tangent bundle  $\Cal{T}\cL$ carries a natural structure of a bundle of Malcev algebras $(\Cal{T}_e \cL, [\cdot, \cdot])$,
that describe the  ``first order approximation" to the Moufang composition law $*$.}

\medskip
 
 For a proof, see the first pages of [Sa61], or [Pa03]. 
 In the case of a CML the tangent Malcev algebra is an abelian Lie algebra.

\bigskip

\centerline{\bf 3. Symmetries of cones of probability distributions}

\smallskip 
\centerline{\bf  on finite sets}

\medskip

{\bf 3.1. Spaces of probability distributions.} The setup on which we focus our attention now is briefly described
in Sec. 1 and 4 of [CoMaMar21] (see also [CoMa20]). Much more details a reader can find in the
primary sources [Vi63] and [Mar19]. See also [MoChe91].

\smallskip

Briefly, let $X$ be a finite set, and $R^X$ the space of  real--valued functions
$X\to \bold{R}$. A classical probability distribution on $X$ consists
of functions $X\ni x\mapsto p_x \in \bold{R}$ such that all $p_x$ are non--negative, and
$\sum_x p_x = 1.$ Thus, the space of such distributions is a simplex $\Delta_X$ of dimension
$\roman{card}\, X - 1$, with the set of vertices that can be canonically identified
with $X$. It is convenient also to consider the open simplex ${}^{\circ}\Delta_X$
consisting of points $(p_x)$ with all $p_x>0$.

\smallskip

 ${}^{\circ}\Delta_X$ is a convex domain in $R^X$, if $\roman{card}\,X \ge 2$.
 Here we understand convex domains in the sense of the Def. 7
in Ch. I, Sec. 5 of [Vi63].

\smallskip

We will also consider the family of functions $X\ni x\mapsto q_x$, with $q_x >0$ for all $x$.
This is a convex cone in $R^X$ in the sense of  Def.1 in Ch. 1, Sec. 1 of [Vi63].
This is the {\it cone fitted onto the convex domain ${}^{\circ} \Delta_X$}
in the sense of Def. 9 in Ch. I, Sec. 5 of [Vi63].

\medskip

{\bf 3.2. Symmetries.} Let $S_X$ be the group of all permutations
of $X$. We will write the left action $S_X\times X \to X$ as
$(s, x) \mapsto s(x)$. By linearity, it extends to the left action
$S_X\times {}^{\circ} \Delta_X \to {}^{\circ} \Delta_X$.

\smallskip

The central construction of [Vi63], surveyed also in [CoMa20],  [CoMa21], establishes
that this space carries a family of canonical Riemannian metrics. One element of such a family
is defined, for example, by a choice of its {\it center}: a  point $c\in {}^{\circ} \Delta_X$. 
 As soon as $c$ is chosen, the geodesics with respect to this metric
are segments of real lines in $\Delta_X$ containing $c$ and one of the vertices $x$.
It is important to keep in mind that the distance from $c$ to the intersection point
of such a real line with interior part of the boundary face of $\Delta_X$ is
infinite.

\smallskip

Moreover, each such geodesic then defines an automorphism of order two $t_x$ of 
metric space ${}^{\circ}\Delta_X$, such that $t_x(c)=c$ and the respective geodesic is
$t_x$--invariant. This action naturally extends to $\Delta_X$.

\smallskip

However, action of $t_x$ on other vertices induces generally a non--trivial
permutation of them.

\medskip

{\bf 3.3. Proposition.} {\it The family of maps above satisfies identities
$$
(t_x t_y t_z)^2 = id \eqno(3.1)
$$
where $id$ is the identical map $X\to X$.}  

\medskip

{\bf Proof.} Intuitively, the symmetries $t_x$ for $x\in X$ can be considered
as {\it boundary limits}  of involutions $s_c$, defined for every point 
$c \in {}^{\circ} \Delta_X$.  Along any geodesic, passing through $c$,
this involution acts a "mirror symmetry": it preserves the Riemannian distance
between $c$ and a variable point $d$, but reverses the direction from $c$ to
$d$, so that $s_c^2= id$.
\smallskip

We will skip here an easy reasoning, showing that this intuition
works, and relations (3.1) indeed follow from relations 
$$
(s_cs_ds_e)^2 = id  \eqno(3.2) 
$$
\smallskip

The relations (3.2) themselves constitute the content of
a classification theory: see [Ma86], Ch. 1, and
more recent publications [Lo69], [SpVe00].
$\blacksquare$

\medskip

Now consider  a general (pseudo)--Riemannian manifold $M$
endowed with a family of isometric involutions $s_c$, $c\in M$, satisfying
relations (3.2).

\smallskip

Define the multiplication law $*: M\times M \to M$ by
$$
c*d := s_c(d)  \eqno(3.3)
$$

\medskip

{\bf 3.4. Theorem.} {\it The set of points of $M$ endowed with
composition law $*$ is  an abelian symmetric quasigroup.} 

\bigskip

The proof is rather straightforward, and we omit it.

\medskip

This statement  can be considered as a bridge between Moufang loops
and symmetries of spaces of probability distributions. In fact, as is shown
in [Ma86], the following family of identities holds. Let $(E,*)$ be a CML, $c\in Z(E)$.
Define maps $t_x: E\to E$ by $t_x(y) :=cx^{-1} *y^{-1}$. These maps satisfy relations (2.1).


\bigskip
\bigskip

\centerline{\bf 4. $F$--manifolds, 3--webs, and Malcev algebras}

\medskip

{\bf 4.1. $F$--identity.} Consider a linear space (or a sheaf of linear spaces)
$\Cal{T}$, endowed with two bilinear operations: commutative and associative
binary composition $\circ$ and Lie bracket $[,]$. 

\smallskip

Define the Poisson tensor $P: \Cal{T} \otimes \Cal{T}\otimes \Cal{T} \to \Cal{T}$
by
$$
P_X(Z,W) := [X,Z\circ W] - [X,Z]\circ W - Z\circ [X,W].    
$$
By definition, $F$--identity is the following constraint upon $(\circ ,[,] )$ ([CoMa20]):
$$
P_{X\circ Y} (Z,W) = X\circ P_Y(Z,W) + Y\circ P_X(Z,W) . \eqno(4.1)
$$
\smallskip

Here we would like to understand a connection between $F$--identity
and main identities defining Malcev algebras.

\smallskip

From the first sight, Malcev's identities differ from  $F$--identity:
they impose linear relations upon operadic monomials of two and three
variables (cf. (2.1)), whereas (4.1) consists of operadic
monomials in four variables.  Following [Na88] and [Na92] we will show,
how to overcome this obstacle.

\medskip

{\bf 4.2. Differential 3--webs.}  Let $M$ be a smooth manifold of dimension $2r$, ($r\ge 1$),
$\Cal{T}M$ its tangent bundle.  A (differentiable) {\it 3--web
on $M$}  is defined as a family of three foliations on $M$ of rank $r$, encoded
by their tangent subbundles:
\smallskip
\centerline{\it horizontal\  tangents\ $\Cal{T}^hM \subset \Cal{T}M$,}
\smallskip

\centerline{\it vertical\ tangents\ $\Cal{T}^vM \subset \Cal{T}M$,}

\smallskip
\centerline{\it transversal\ tangents\  $\Cal{T}^tM \subset \Cal{T}M$,}
\smallskip
such that {\it the direct sum of any two different members of this family
coincides with $\Cal{T}M$.}

\medskip

As in [Na88], denote by $H$, resp. $V$, $T$, the projection operator 
$\Cal{T}M\to \Cal{T}M$ with kernel $\Cal{T}^vM + \Cal{T}^tM$, resp. $\Cal{T}^hM + \Cal{T}^tM$,
$\Cal{T}^hM + \Cal{T}^vM$.

\smallskip

From the definition, it follows that $H^2=H, V^2 = V$, $T^2=T$.
Moreover, one easily sees that there exists a unique  operator
$J$ on $\Cal{T}M$ such that $J^2=Id$, $HJ + JH =J$,
and $J$ induces an isomorphism between $\Cal{T}^hM$  and $\Cal{T}^vM$.

\smallskip

Such a pair is called an $\{H,J\}$--structure on $M$, and it carries
exactly the same information as needed for a description of a 3--web on $M$. 
The following results (Theorems 3.2 and 3.4 in [Na88]) play the crucial role
in the following:

\medskip

{\bf 4.2.1. Proposition.}  {\it (i) Consider a manifold $M$ with an $\{H,J\}$--structure.
Then there exists a unique covariant derivation $\nabla$ on $M$ such that
$\nabla H =\nabla J = 0$, and that the torsion tensor
$t(X,Y) := \nabla_{HX}Y - \nabla_{VY}X -[HX,HY] = 0$ for any $X,Y$.

\smallskip

(ii) This $\{H,J\}$--structure comes from a 3--web, that is, respective distributions
are integrable, if and only if the following conditions are satisfied:
$$
Vt(HX,HY) = Ht(VX,VY) = 0, 
$$
$$
JHt(HX,HY) +Vt(JHX, JHY) =0. \eqno(4.2)
$$
}

\medskip

We do not reproduce the proof here.

\medskip

The covariant derivative $\nabla$ defined in Prop. 4.2.1, is called the 
{\it canonical connection}, or {\it Chern connection} of the respective 
$\{H,J\}$--structure.

\medskip

Let now $M$ be a space of probability distributions. As was shown
in Sec. 3, it has a family of structures of CMLs, and thus of quasigroups.
The latter one produces a family of (bundles) of Malcev algebras.

\medskip

{\bf 4.3. Theorem.} {\it In this setup, $M$ admits a family of 3--webs,
whose Chern connections define compatible flat structures 
of the respective family of $F$--structures on $M$, in the sense
of [CoMa20], Sec. 2.3.}

\medskip

A key observation proving this is the formula $X\circ Y= [X, [Y,C]]$ 
in Sec. 2.3 of [CoMa20], its comparison with (4.2) above.
For additional information, see [ASh92], [CoCoNen21].

\bigskip

\centerline{\bf 5. General Moufang loops and codes}

\medskip

{\bf 5.1. Non(necessarily)commutative Moufang loops (ML).} We keep
the notation $*$ for binary multiplication, but add or change
other essential notations and conditions (cf. [Gri86] and references therein).

\smallskip

{\bf 5.1.1. Definition.} {\it (i) A loop $\Cal{L}$ is a set with a binary composition law
$*\,:\,  \Cal{L} \times \Cal{L} \to \Cal{L}, (a,b) \mapsto a*b$, endowed by
two--sided unit, denoted $1$ if it will not lead to confusion, and such that for each elements  $a,b\in \Cal{L}$ 
the equations $a * x =b$ and $y * a =b$ have a unique solution in $\Cal{L}$, denoted $a^{-1}$.

\smallskip
(ii) A loop $(\Cal{L}\,, *\,)$ is called Moufang, if any triple
$(a,b,c )\in \Cal{L}^3$ satisfies the ``near--associativity" relation
$$
(a*b)*(c*a) = a*((b*c)*a) .
$$

\smallskip

(iii)  The operations  commutator $[a,b]$ and  associator $[a,b,c]$
in a Moufang loop $\Cal{L}$ are defined as follows:
$$
[a,b] := (a*b)*((b*a)^{-1}) , \quad [a,b,c] := ((a*b)*c) *((a*(b*c))^{-1}) .
$$
}
\smallskip

{\bf 5.1.2. Definition.} {\it Let $(\Cal{L},*)$ be a Moufang loop.

\smallskip

(i) The Moufang centre $C(\Cal{L})$ is the set of all elements $a\in \Cal{L}$ such that 
$[a,b]=1$ for each $b \in \Cal{L}$.

\smallskip

(ii) The  nucleus $N(\Cal{L})$ of  $\Cal{L}$ is the set of all elements $a\in \Cal{L}$
such that for any $b,c\in \Cal{L}$ we have
$$
[a,b,c] = [b,a,c] = [b,c,a] =1.
$$

(iii) The centre $Z(\Cal{L})$ is defined as $N(\Cal{L})\cap C(\Cal{L})$.
}

\smallskip

One can easily check that  the nucleus $N(\Cal{L})$ is a subgroup of $\Cal{L}$, and the centre
$Z(\Cal{L})$ is an abelian subgroup.

\smallskip

Let now $p$ be a prime.
We denote by $\Cal{L}_p$ the set of elements of $\Cal{L}$
whose order is a power of $p$. The torsion subloop of a Moufang loop $\Cal{L}$ is the direct product of the $\Cal{L}_p$
over all primes $p$ (see [Hsu00a]). It is also shown in [Hsu00a] that, if $\Cal{L}$ is a Moufang loop such that 
$\Cal{L}/Z(\Cal{L})$ is an abelian group, then for all $p>3$, the subloop $\Cal{L}_p$ is a group. 
A Moufang loop $\Cal{L}$ is caled a $p$--loop if every element of $\Cal{L}$ has order a power of $p$. 
For a Moufang loop, the order of any element divides the order of the loop, hence 
a Moufang loop of order a power of $p$ is a finite $p$--loop, $\Cal{L}=\Cal{L}_p$, hence 
in particular a group, if $p>3$ (see Theorem~A of [Hsu00a]).

\medskip

{\bf 5.2. Error--correcting codes.} The family of codes with which we deal
in this paper can be described as follows (see [MaMar12] and references therein).

\smallskip

Let $q\ge 2$ be an integer, and $A$ (or $A_q$) a finite set of cardinality $q$ (alphabet).
A sequence $(\alpha_i)$ of elements of $A$, $i=1,2,\dots , n$, is called  a {\it word
of length $n$}. By definition, a code $C$ is a non--empty subset $C\subset A^n$.

\smallskip

Define the  {\it Hamming distance} between two words $(\alpha_i)$ and $(\alpha_i^{\prime}$ of
the same length as 
$$
d((\alpha_i),(\alpha_i^{\prime})) := \roman{card}\, \{ i \in (1, \dots , n)\,|\, \alpha_i \ne \alpha_i^{\prime} \}.
$$
\medskip

Given such a code $C$, we denote by $n(C)$ the common length of all words in $C$,
by $d(C)$ the minimal distance between two different words in $C$, and by 
$k(C)$ the number $[\roman{log}_q\roman{card}\,(C)]$. The quadruple of
integers $[n,k,d]_q$ defines a finite family of codes $C\subset A_q^n$, for which
$d=d(C)$ and $k=k(C)$.

\smallskip

A code $C$ becomes a kind of ``dictionary of an artificial language'' as soon as
one ascribes to words in $C$ some meanings, ``semantics''. Finite sequences of code words
are ``sentences''.

\smallskip

If then an information encoded by such a sentence ought to be transmitted, say, by broadcasting, it might
become distorted. The idea of error--correcting codes consists in imagining
that noise in such a channel, with large probability, distorts only rare letters
in code words. Hence, if the distance between two different code words is big enough,
one can recognise the distorted letters and to correct them. We must pay for it
by using code words of larger length that is strictly necessary for encoding
relevant information.

\smallskip

For this reason the number 
$R(C) := k(C)/n(C)$ is called the {\it transmission rate} of $C$,
and the number $\delta (C) := d(C)/n(C)$ is called the {\it relative
minimal distance} (between code words) of $C$.

\smallskip

The words ``geometry of information'' in this setup refer
to the geometry of the set of {\it code points} $P_C := (R(C), \delta (C)) \in [0,1]^2$.

\smallskip

{\bf 5.2.1. Unstructured vs structured codes.} If sets of code words
are endowed with additional
data/restrictions,  we call generally the respective  $A_q$--codes 
``structured'' ones.

\smallskip

 Two most studied classes of structured codes are the following ones:
  
  \smallskip
  
 (i) {\it Linear codes}. Here $A_q := \bold{F}_q$, finite field
 of cardinality $q$, and $C\subset \bold{F}_q^n$ are $\bold{F}_q$--linear subspaces.
 
 \smallskip
 
 (ii) {\it Algebraic--geometric codes.} For the same class of alphabets, one can consider
$\bold{F}_q$--points in affine (or projective) $\bold{F}_q$--schemes with a chosen
coordinate system.
 
 \smallskip
 
 As we will see below, Moufang symmetries generally become visible in
 special structured codes. See also an unusual setup of [PeSuWeiZa20].

\medskip

{\bf 5.3. Code loops.} 
Code loops were originally constructed (see [Gri86] and also [Na08], [NaRob21])
from a family of $\bold{F}_2$--linear codes (including the Golay code), as a generalization of
extensions given by cocycles.  In
this setting, one considers a linear code $C\subset \F_2^n$ and a function
$\theta: C \times C \to \F_2$.  If the function $\theta$ satisfies the cocycle identity
$$ 
\theta(v,w)-\theta(u+v,w) + \theta(u, v+w) - \theta(u,v) =0, 
$$
then one obtains the Heisenberg group $H(C,\theta):=C\ltimes_\theta \F_2$, with multiplication
$$
 (v,x)\star_\theta (w,y)=(v+w, x+y+\theta(v,w)). 
$$
(see further details below). This multiplication is associative, but generally
noncommutative.

\smallskip

In order to remake $C\ltimes_\theta \F_2$ into a nonassociative loop with identity element $(0,0)$,
one replaces the cocycle identity by the {\it twisted cocycle identity}
$$ 
\theta(v,w)-\theta(u+v,w) + \theta(u, v+w) - \theta(u,v) = \delta(u,v,w), 
$$
where twisting $\delta: C\times C \times C \to \F_2$ is a certain function.

\smallskip

Theorem~A of [Hsu00a] implies that, for any $p>3$, Moufang $p$--loops $\cL$ (defined as recalled  
at the end of Section~5.2.1) that are obtained as central extensions of
a code $C\subset \F_q^n$, with $q=p^r$, by the center $Z(\cL)=\F_q$ are in fact {\it groups}. 
Thus, all these cases fall within the framework of usual construction of central extensions of groups.
in Section~5.1. However, for $p=2$ and
$p=3$ one has interesting non-associative code loops. To better compare these cases to
the setting of Section~5.1, we will recall, at the beginning of Sec. 6,
the general construction of [Hsu00a], [Hsu00b]
of Moufang loops obtained as central extensions of Frattini type, before introducing
our new, more general construction in Sec. 6.5. 

\medskip

{\bf 5.4. Geometry of information: asymptotic bounds for error--correcting codes.} 
We return here to the definition of code points at the end of subsection 5.2 above.
Fix cardinality $q$ of an alphabet, and consider a class of error--correcting codes $Cod_q$ 
with this alphabet. Denote by $cp$ the map $Cod_q \to [0,1]^2$ sending $C\in Cod_q$ to
$P_C$. The {\it multiplicity} of a code point $x$ is defined as the cardinality of $cp^{-1}(x)$.

\smallskip

{\bf 5.4.1. Definition.} {\it A continuous function $\alpha_q(\delta), \delta \in [0,1]$ is  called 
the  asymptotic bound for the family $Cod_q$, if it satisfies the following conditions:

\smallskip

(i) The set of code points of infinite multiplicity is exactly the set of rational points $(R,\delta )$
satisfying $R\le \alpha_q (\delta )$.

\smallskip

(ii) Code points of finite multiplicity all lie above the asymptotic bound and all are isolated:
a sufficiently small open neighbourhood of each such point contains no other code points.}

\medskip

{\bf 5.4.2. Theorem.}  {\it Asymptotic bounds exist 
 \smallskip
 
 (i) For unstructured codes.
 
 \smallskip
 
 (ii) For linear codes over $\bold{F}_q$ where $q$ is a power of prime.
 
 \smallskip
 
 (iii) For certain classes of algebraic--geometric codes over finite fields.}
 
 \medskip

In its present final form this theorem was proved in [Ma12].

\medskip

Moufang symmetries appear not directly in this geometric picture, but rather
in various formalisms motivated by theoretical physics and
based on the vision of an asymptotic bound as a phase transition
curve, in a classical or quantum version.

\bigskip

\centerline{\bf 6. Moufang loops, almost symplectic structures,}
\smallskip

\centerline{\bf and quantum codes}

\medskip

In the papers [CRSS97], [CRSS98] an algorithm was introduced,
producing quantum codes from self--orthogonal classical codes.
We will call it the CRSS algorithm.

\smallskip

In [HMPS18] the CRSS algorithm associating quantum codes to self--orthogonal
classical codes was reformulated geometrically in terms of the canonical
quantisation of symplectic spaces over finite fields of [GuHa09], via representations of
Heisenberg groups. 

\smallskip

The main result of this section is a new construction of code loops
based on canonical symplectic quantization over finite fields, adapted to an
almost--symplectic case. We also show that the code loops obtained in this way
have associated CRSS quantum codes determined by isotropic subspaces,
and perfect tensors associated to Lagrangians. 

\smallskip

To compare our approach with previous constructions of code loops, it is
important to note the following.
The code loops described in [Gri86], as well as the more general construction 
in [Hsu00a], [Hsu00b] of Moufang loops $\cL$ that are central extensions
$$ 0 \to Z \to \cL \to C \to 0, $$
of abelian groups $Z=Z(\cL)$ and $C=\cL/Z(\cL)$, rely on introducing 
a notion of ``cubic symplectic structure" (see [Hsu00a], [Hsu00b] for more details).
In the case where $Z(\cL)\simeq \Z/p\Z$, Moufang
loops $\cL$ obtained in this way are called {\it Frattini extensions}. 

\smallskip

Results of [Hsu00a] on the Moufang $p$--loops 
imply that, for $p>3$, loops $\cL$ obtained as central extensions of
a code $C\subset \F_p^n$ by $Z(\cL)=\F_p$ are in fact always groups. 

\smallskip

However, for $p=2$ and $p=3$ one has interesting non--associative code loops. 
In the case of $p=2$ it follows from [Hsu00a], [Hsu00b], and [ChGo90] that
all the code loops obtained through the ``cubic symplectic structures" of [Hsu00a]
can be realised by doubly even codes as in the construction of [Gri86]. This means that
$C$ is a binary linear code $C\subset \F_2^n$ that is {\it doubly even}, namely
the weight $|v|=\# \{ v_i=1 \}=v_1+\cdots+v_n$ of the code words (the number of 
ones in the word) is divisible by $4$, and the twisted cocycle $\theta$ that
gives the code loop (see Section~5.3 above) has twisting function
$$ 
\delta(u,v,w) = | u \& v \& w |\mod 2, 
$$
where $u \& v:=(u_1 v_1, \ldots, u_n v_n)$ denotes the logical AND operation, with
$$ 
\theta(v,w)+\theta (w,v)=\frac{1}{2} | v \& w| \mod 2
$$ 
$$
 \theta(v,v)=\frac{1}{4} |v| \mod 2. 
 $$

\smallskip

In the construction we present in this section, the twisted cocycle is an
almost-symplectic structure (as we discuss more precisely below). For $p>2$,
our construction is a special case of the central extensions mentioned above,
but in the case $p=2$ they are different, following the setting of [GuHa12] 
for symplectic quantization in characteristic $2$. 

\medskip

{\bf 6.0.1. Question.} Let $\cL$ be a code Moufang loop corresponding to a
doubly even code $V$. Describe the set $\roman{Cod}(\cL)$ of all doubly 
even codes $W$ such that the corresponding Moufang loop is isomorphic to $\cL$.

\medskip

{\bf 6.1. Quantum codes.} As in 5.2.1 above, consider an alphabet of cardinality
$q=p^r$ endowed with a structure of $\F_q$--linear space of dimension $n$.
Call $\cV_q=\C^q$ a {\it single $q$--ary qubit} space. In the case $q=2$ we refer to it just as 
the {\it single qubit space}. Then $\cV_q^{\otimes n}$ is the space of $n$ $q$--ary qubits. A {\it quantum
error} on the vector space $\cV_q^{\otimes n}$ is a linear operator $E$ of the form 
$E=E_1\otimes \cdots \otimes E_n$.

\medskip

{\bf 6.1.1. Definition.} {\it Consider on the space $\cV_q^n=(\C^q)^{\otimes n}$
of $n$ $q$--ary qubits the orthonormal basis $| a \rangle$, parameterized by vectors $a=(a_1,\ldots, a_n)\in \F_q^n$.
On $\cV_p=\C^p$ the bit flip and phase flip operators $T$ and $R$ are defined in this basis as
$$
 T=\pmatrix 0 & 1 & 0 & \cdots & 0 \\ 0 & 0 & 1 & \cdots & 0 \\ \vdots & & & \ddots & \vdots \\
0 & 0 & 0 & \cdots & 1 \\
1 & 0 & 0 & \cdots & 0  \endpmatrix ,\ \ \  R = \pmatrix
1 &  & & & \\     & \xi & & & \\   &  & \xi^2 & & \\   &  &  & \ddots & \\   & & & & \xi^{p-1} 
\endpmatrix ,
$$
where $\xi$ a $p$--th root of unity, $\xi^p=1$. 

\smallskip

For any pair of vectors 
$a=(a_1,\ldots, a_n), b=(b_1\ldots,b_n)\in \F_q^n$, the error operators $E_{a,b}$ on
the space $\cV_q^n=(\C^q)^{\otimes n}$ is defined in the above basis as
$$ 
E_{a,b}=T_a R_b = (T_{a_1} \otimes \cdots \otimes T_{a_n}) (R_{b_1}\otimes \cdots \otimes R_{b_n}), 
$$
for $a=(a_1,\ldots, a_n), b=(b_1\ldots,b_n)\in \F_q^n$, with 
$$ 
T_{a_i} :=T^{a_{i1}}\otimes \cdots \otimes T^{a_{ir}}, \ \ \ R_{b_j} := R^{b_{j1}}\otimes \cdots \otimes R^{b_{jr}}.
$$

Here we look at  $\F_q$ as an $r$--dimensional vector space over $\F_p$, with
$a_i=(a_{i\ell})_{\ell=1}^r$ and $b_j=(b_{j,\ell})_{\ell=1}^r$, with entries $a_{i\ell}, b_{j,\ell}\in \{0, \ldots, p-1\}$
in the exponents of the corresponding powers of the bit flip and phase flip operators $T$ and $R$.
}

\medskip

The bit flip and phase flip operators satisfy $T^p=R^p=\roman{id}$ and the commutation relation $TR =\xi RT$.
The error operators $E_{a,b}$ define a linear basis of $M_{q^n\times q^n}(\C)$, orthonormal with respect 
to the inner product $\langle A, B\rangle=\Tr(A^* B)$.  
In particular, they generate all possible quantum errors on $\cV^{\otimes n}$. 

\medskip
  
An $[[n,k,d]]_q$ {\it quantum error--correcting code} is a subspace $\cC \subset \cV^{\otimes n}$, 
spanned by vectors $|a\rangle$ with $a$ in a subspace of dimension $k$ over $\F_q$. 
So it can correct $\le d-1$ errors. This means that, 
for every error operator of the form $E=E_1\otimes \cdots \otimes E_n$, with
$\omega(E)=\roman{card} \{ i\,:\, E_i\neq I \}< d$, the orthogonal projection  $P_\cC$ onto 
$\cC$ in $\cV^{\otimes n}$ satisfies 
$$ 
P_\cC E P_\cC =\lambda_E \, P_\cC . 
$$
The definition of quantum error--correcting codes in 
 [CRSS97], [CRSS98] describes the same family of codes:

\medskip

{\bf 6.1.2. Definition.}{\it 
A quantum error--correcting code is a subspace $\cC \subset \cV^{\otimes n}$
given by a joint eigenspace of the operators $E_{a,b}$ in an abelian subgroup $\cS$ of the group
$\cG_n = \{ \xi^i E_{a,b}, \, a,b\in \F_q^n, \, 0\leq i \leq p-1 \}$. } 

\medskip

{\bf 6.2. The symplectic CRSS algorithm.} The CRSS (Calderbank--Rains--Shor--Sloane)
algorithm of [CRSS97], [CRSS98] constructs quantum codes from classical self--orthogonal 
error correcting codes. In [HMPS18] this construction was generalised using symplectic
quantisation over finite fields, so that the original self--orthogonal case is recovered as a
special case of this geometric construction. We refer to this version as the {\it symplectic CRSS algorithm}.
We recall here briefly the results of Sections~2 and 3 of [HMPS18] where this construction is
presented.

\smallskip

The construction of [HMPS18] relies on the functorial geometric quantisation 
of symplectic vector spaces over a finite field (of characteristic $p>2$)
developed in [GuHa09].  

\smallskip

{\bf 6.2.1. Definition.} {\it Let $q=p^r$ with $p$ odd. A symplectic vector space 
$(V,\omega)$ consists of a finite dimensional vector space $(V,\omega)$ of 
dimension $2n$ over $k=\F_q$, together with a symplectic form $\omega$,
namely a function  $\omega: V\times V \to k$ that is antisymmetric $\omega(u,v)=-\omega(v,u)$
and non-degenerate, namely for all $u\neq 0$ in $V$ there exists a $v\in V$ with $\omega(u,v)=1$,
and that is moreover closed, namely it satisfies the cocycle condition
$$ 
d\omega(u,v,w)=\omega(v,w)-\omega(u+v,w)+\omega(u,v+w)-\omega(u,v) =0. 
$$ 
The Heisenberg group $\roman{Heis}(V,\omega)$ is the central extension
$$ 
0 \to k \to \roman{Heis}(V,\omega) \to V \to 0 
$$
determined by the cocycle $\omega$. }

\medskip

The multiplication in $\roman{Heis}(V,\omega)$ is given by
$$ 
(v,x)\cdot (w,y) =(v+w,x+y+\frac{1}{2} \omega(v,w)). 
$$
The cocycle condition ensures the associativity of this multiplication law. The center
of the Heisenberg group is $Z(\roman{Heis}(V,\omega))=\{ (0,x)\,:\, x\in k\}$.
The cocycle condition $d\omega=0$ is the vanishing of the Hochschild differential.

\smallskip

The choice of a central 
character $\chi: k \to \C^*$ determines an irreducible complex representation $\cH_\chi$ of the Heisenberg group
$\roman{Heis}(V,\omega)$, the {\it Heisenberg representation $\pi_\chi$.}

\smallskip
 
 As described in [GuHa09],
the choice of a Lagrangian subspace $L \subset V$ determines a model for the Heisenberg representation
$$ 
\pi_{L,\chi}: \roman{Heis}(V,\omega) \to \GL (\cH_{(V,L,\omega,\chi)}), 
$$
where $\cH_{(V,L,\omega,\chi)}$ is the subspace of the space $\C[\roman{Heis}(V,\omega)]$ of complex
valued functions on the set $\roman{Heis}(V,\omega) \simeq V \times k$ that satisfy
$$
 f((0,x)\cdot (w,y))=\chi(x)\, f(v,y), \ \ \forall x\in k, \,\, \forall (w,y)\in V \times k,
 $$
$$ 
f((v,0)\cdot (w,y))= f(w,y), \ \ \forall (v,0)\in L, \,\, \forall (w,y)\in V \times k.  
$$
Here $\Heis(V,\omega)$ acts upon on $\cH_{(V,L,\omega,\chi)}$  by right translations:
$$ 
(\pi_{L,\chi}(v,x)\, f)(w,y)= f((w,y)\cdot (v,x)). 
$$

\smallskip

This version is modelled on
the usual construction of the quantum mechanical Hilbert space that identifies the position 
and momentum representations with a Lagrangian subspace $L$ and its dual space $L^\vee$. 

\smallskip

The further enrichment adds to this an {\it orientation on the Lagrangian},  replacing $L$ with a pair $L^o=(L,o_L)$
of a Lagrangian subspace $L \subset V$ and a non--zero vector $o_L\in \Lambda^{top} L$. It determines
{\it intertwining isomorphisms} $T_{L_1^o,L_2^o}: \cH_{(V,L_1,\omega,\chi)} \to \cH_{(V,L_2,\omega,\chi)}$ 
producing a trivialisation of the bundle of the Heisenberg representation models $\cH_{(V,L,\omega,\chi)}$
over the space of oriented Lagrangians. This expresses the functoriality of the geometric
quantisation of  [GuHa09]. 

\smallskip

The symplectic form $\omega$ determines a Darboux decomposition of the
symplectic space $V$ into a sum of $\F_q^2$ subspaces, which in turn, by functoriality
of the geometric quantisation, determines a tensor
product decomposition of $\cH_\chi$ into copies of the space $\C^q$ of a single $q$--ary qubit. 

\medskip

{\bf 6.2.2. Example.} 
In the case of $(\F^{2n}_q,\omega)$ with $\omega$ the standard Darboux form, 
the Heisenberg group $\Heis(\F^{2n}_q,\omega)$ representation with the central
character determined by a $p$--th root of unity $\xi$ with $\xi^p=1$, is given by
the error operators $E_{ab}=T_a R_b$ of Definition~6.1.1.

\medskip

This example is the key to the relation between Heisenberg group representations
(that is, the functorial geometric quantization of [GuHa09]) and the construction of
quantum error correcting codes. The main result is summarized as follows.

\medskip

{\bf 6.2.3. Proposition  ([HMPS18]). }  {\it Let $(V,\omega)$ be a $2n$--dimensional
symplectic vector space over $\F_q$ with $q$ odd. An {\it isotropic subspace} $C \subset V$
of dimension $k$ determines an abelian subgroup of the Heisenberg group 
$\Heis(V,\omega)$, that is, mutually diagonalisable error operators
in the corresponding representation $\cH_\chi(V,\omega)$.  
Each joint eigenspace of $C$ in $\cH_\chi(V,\omega)$ gives a quantum code $\cC_C$
associated to the classical code $C$.}

\medskip

We refer to the assignment $C\mapsto \cC_C$ as the symplectic CRSS algorithm. 
The original construction of [CRSS97], [CRSS98] for self--orthogonal classical codes is 
a special case of this symplectic construction, see Sections~2 and 3 of [HMPS18] for 
more details. 

\medskip

{\bf 6.3. Perfect tensors.}  Let $\cV=\C^q$ be the single $q$--ary qubit space. An $m$-tensor
is an element $T\in \cV^{\otimes m}$. We write such a tensor, in the standard
basis $\{ |a\rangle \}_{a=(a_1,\ldots, a_m)\in \F_q^m}$ of Definition~6.1.1, 
as $T=(T_{a_1,\ldots, a_m})$ with $m$ indices $a_i$. We assume that $\cV$ is endowed 
with an inner product to identify it with its dual. This means that we can raise and lower indices
of $T$: after raising $j$ indices we can identify $T$ with an element in 
$Hom(\cV^{\otimes j}, \cV^{\otimes (m-j)})$. We refer to such an identification as a $(j,m-j)$-splitting 
(bipartition) of the indices of the tensor $T$.

\smallskip

{\bf 6.3.1. Definition.} {\it A {\it perfect tensor} $T$ is a tensor in $\cV^{\otimes m}$,
such that, for any $j\leq m/2$, all resulting splittings of the set of indices are {\it isometries}
$$ T: \cV^{\otimes j} \to \cV^{\otimes (m-j)}. $$}

\smallskip

A perfect tensor determines a perfect code that encodes one $q$-ary qubit to $m-1$ $q$-ary qubits.
These quantum codes realize maximal entanglement across bipartitions. 
Tensor networks obtained by contracting legs of an arrangement of perfect tensors along
a tessellation of a hyperbolic space have been considered in the context of the AdS/CFT 
holographic correspondence in string theory as discretizations of the bulk geometry that
produce entangled boundary states, in such a way that the entanglement entropy on the
boundary is expressible in terms of geodesic lengths in the bulk, according to the
Ryu--Takayanagi conjecture. We refer the reader to [PYHP15] for this role of perfect
tensors and tensor networks. This has become a prominent area of research in 
models of AdS/CFT holography.

\smallskip

Here we just recall a result of [HMPS18] that shows how perfect tensors can be
constructed from the geometric quantization of [GuHa09] of symplectic spaces over finite fields
of characteristic $p>2$, through the geometry of Lagrangian subspaces.

\smallskip

Consider a symplectic vector space $(V,\omega)$ over $\F_q$ and a 
Lagrangian subspace $L\subset V$ (necessarily of dimension $\dim(V)/2$).
Let $V =\oplus_i V_i$ be the Darboux decomposition of $(V,\omega)$, with
$V_i\simeq \F_q^2$, and let  $\cH=\otimes_i \cH_i$ be the corresponding
decomposition of the irreducible Heisenberg representation in $q$--ary qubits. 
A choice of splitting of the indices of a tensor $T$ in $\cH$ corresponds to
a decomposition $V=W\oplus W'$, given by a partition of the  pieces of the 
Darboux decomposition.

\smallskip

We will say that a Lagrangian $L$ is in {\it general position} with 
respect to a decomposition $V=W\oplus W'$ if the intersections $L\cap W$ 
and $L\cap W'$ with the pieces of the decomposition are as small as possible.

\medskip

{\bf 6.3.2. Proposition. ([HMPS18]). }  {\it A Lagrangian $L$ that is in general position with respect 
to the Darboux decomposition $V =\oplus_i V_i$, determines a symplectomorphism
$\psi_L : \bar W \to W'$ for a given splitting $V=W\oplus W'$ as above into half-dimensional pieces, 
with $(\bar W, \bar\omega)=(W,-\omega)$ the dual symplectic space. The
corresponding map $\cH(\psi_L): \cH(W)^\vee \to \cH(W')$ under the quantization
functor of  [GuHa09] is a perfect tensor $$ T_L \in \cH(W)\otimes \cH(W')=\cH(V). $$
}

\medskip

One of our main goals in the rest of this section will be a generalisation of Propositions~6.2.3 and
6.3.2 to code loops.  To this purpose, we first have to discuss the case of characteristic $2$,
which was not considered in [HMPS18], since the most interesting  code loops arise in characteristic $2$.

\medskip

{\bf 6.4. Geometric quantisation in characteristic $2$.} 
The functorial quantisation of symplectic vector spaces over finite fields
requires a separate treatment for the case of characteristic $p=2$, for
which we recall the setting of [GuHa12].  

\smallskip

Consider a finite field $\F_{2^r}$. We will identify it with
residue field $\cO_K/\fm_K=\F_{2^r}$ of an unramified extension $K$ of degree $r$
of $\Q_2$. More precisely, let $\cO_K\subset K$ the ring of integers and $\fm_K$ the maximal
ideal. Consider the ring $R=\cO_K/\fm_K^2$.
Let $(\tilde V,\tilde\omega)$ be a free $R$--module endowed with a symplectic form.
The $\F_{2^r}$--vector space $V=\tilde V/\fm_K$ is endowed with a $R$--valued
non--degenerate skew--symmetric form determined by $\omega=2\tilde\omega$. 
In the following, when we say that $(V,\omega)$ a symplectic vector space over 
$\F_{2^r}$, we mean a pair obtained as described here, with an $R$--valued form $\omega$.

\medskip

{\bf 6.4.1. Definition.} {\it 
A polarization of the symplectic form $\tilde\omega$ is a bilinear form
$\tilde\beta: \tilde V \times \tilde V \to R$ with 
$\tilde\beta(\tilde v,\tilde w)-\tilde\beta(\tilde w, \tilde v)=\tilde\omega(\tilde v, \tilde w)$.}

\medskip

Note that bilinearity implies the cocycle condition
$$ \tilde\beta(\tilde v, \tilde w+\tilde u)- \tilde \beta(\tilde v, \tilde w) -\tilde \beta(\tilde v+ \tilde w, \tilde u)
+ \tilde\beta(\tilde w,\tilde u)=0. $$
Setting $\beta=2\tilde \beta$ induces an $R$--valued cocycle on $V$ with  
$\beta(v,w)-\beta(w, v)=\omega(v, w)$.

\medskip

{\bf 6.4.2. Definition.}{\it 
The Heisenberg group in the characteristic $2$ case is the extension
$$ 0 \to R \to \Heis(V,\beta) \to V \to 0 $$
determined by the cocycle $\beta$ as above, with multiplication
$$ (v,r)\star (w,s)=(r+s+\beta(v,w), v+w). $$}

\medskip

The choice of a character $\chi: R \to \C^*$ determines an irreducible complex representation
$\cH_\chi(V,\beta)$ of $\Heis(V,\beta)$.

\smallskip

Following [GuHa12] we also consider the realisations of this representation,
associated to choices of Lagrangians. Here we need to use its enriched version:
  {\it enhanced Lagrangians}.

\medskip

{\bf 6.4.3. Definition.} {\it Let $(V,\omega)$ be a symplectic vector space over 
$\F_{2^r}$, in the sense described above. An enhanced Lagrangian consists of a pair 
$(L,\alpha)$ where $L\subset V$ is a Lagrangian subspace and $\alpha: L \to R$ satisfies
$$ \alpha(v+w)-\alpha(v)-\alpha(w)=\beta(v,w). $$}

\medskip

This datum $\alpha: L \to R$ defines a section of the projection $\Heis(V,\beta) \to V$
over $L\subset V$ by $\tau :v\mapsto (v,\alpha(v))$ which satisfies $\tau(v+w)= (v+w, \alpha(v+w))=
(v+w,\alpha(v)+\alpha(w)+\beta(v,w))=\tau(v)\star \tau(w)$, for $v,w\in L$. The corresponding
realization $\cH_{(V,L,\beta,\chi)}$ of the Heisenberg representation $\pi_{\chi,L}$ is given by
the subspace of $\C[\Heis(V,\beta)]$ of functions with
$$ f((0,x)\cdot (w,y))=\chi(x)\, f(w,y), \ \ \forall x\in k, \,\, \forall (w,y)\in V \times k, $$
$$ f(\tau(v) \cdot (w,y))= f(w,y), \ \ \forall v \in L, \,\, \forall (w,y)\in V \times k,  $$
with the action of $\Heis(V,\beta)$ by right translations, see [GuHa12]  for more details.

\smallskip

In this case again one can consider an isotropic subspace $C\subset V$. Since $\omega|_C \equiv 0$,
the polarization function $\beta$ restricted to $C$ is symmetric. Given as above a function
$\alpha: C \to R$ with $\alpha(v+w)-\alpha(v)-\alpha(w)=\beta(v,w)$ for all $v,w\in C$, the section
$\tau: C \to \Heis(V,\beta)$, $\tau(v)=(v,\alpha(v))$, determines an abelian subgroup of 
$\Heis(V,\beta)$, since $\beta(v,w)=\beta(w,v)$ on $C$. 

\medskip

Proposition~6.2.3 admits then the following version in characteristic $2$ case.

\smallskip

{\bf 6.4.4. Proposition.} {\it Let $(C,\alpha)$ be a pair of an isotropic subspace of $(V,\omega)$
and an enhancement function $\alpha: C \to R$ satisfying  
$$ \alpha(v+w)-\alpha(v)-\alpha(w)=\beta(v,w), $$
for $\beta$ a polarization of $\omega$. 
A common eigenspace of all the operators $\pi_{\chi,L}(\tau(C))$ on the space $\cH_{(V,L,\beta,\chi)}$
defines a quantum error--correcting code $\cC_{C,\alpha} \subset \cH_{(V,L,\beta,\chi)}$. 
The assignment $(C,\alpha)\mapsto \cC_{(C,\alpha)}$ gives the symplectic CRSS
algorithm for $p=2$. }

\medskip

In the following subsection we use this setting to obtain a new construction of code
loops, given by extensions
$$ 0 \to R \to \cL \to C \to 0, $$
where $C\subset \F_{2^r}^n$ is a linear code endowed with an
almost--symplectic structure, and $R=\cO_K/\fm_K^2$ as above. 

\medskip

{\bf 6.5. Code loops and almost symplectic structures.} We now pass from the
setting of Heisenberg groups to that of code loops by replacing symplectic
structures with almost--symplectic structures.  Our code loops are a direct
natural generalisation of Heisenberg groups, when the symplectic form is
no longer required to be closed and is therefore replaced by an almost--symplectic form.

\medskip

{\bf 6.5.1. Definition.} {\it 
An almost symplectic structure on a finite dimensional vector space $V$ over $\F_q$, with $q$ odd,
is a non-degenerate skew-symmetric form $\omega: V\times V \to \F_q$. Namely $\omega$ satisfies

(i) $\omega(u,v)=-\omega(v,u)$, with $\omega(u,0)=\omega(0,u)=0$, 

(ii) for any $u\neq 0$ in $V$, there is some $v\in V$ satisfying $\omega(u,v)\neq 0$.

The form $\omega$ is not required to be closed and has a nontrivial
coboundary $d\omega=\delta$, given by
$$ d\omega(u,v,w)=\omega(v,w)-\omega(u+v,w)+\omega(u,v+w)-\omega(u,v) =\delta(u,v,w). $$
}

\smallskip

The nontrivial Hochschild coboundary $d\omega=\delta$ is exactly what 
in the literature on code loops is usually referred to as the ``twisted cocycle" condition,
with ``twisting" $\delta$ (see Section~5.3 above). We prefer to use here the coboundary
terminology for consistency with the usual case of almost--symplectic structures 
on manifolds.

\smallskip

We focus here especially on the case of characteristic $p=2$. In this case, we
proceed as in the symplectic case of [GuHa12] recalled above. 

\medskip

{\bf 6.5.2. Definition.} {\it For $q=2^r$, consider as above the ring $R=\cO_K/\fm_K^2$,
where $\F_{2^r}=\cO_K/\fm_K$. An almost--symplectic vector space
$(V,\omega)$ is defined as in Definition~6.5.1, with the almost--symplectic form
$\omega: V \times V \to R$. A polarisation of the almost--symplectic
form is a function $\beta : V \times V \to R$ satisfying the relation
$$ 
\beta(u,v)-\beta(v,u)=\omega(u,v), 
$$
with
$$
 d\beta (u,v,w)=\beta(v,w)-\beta(u+v,w)+\beta(u,v+w)-\beta(u,v) =\gamma(u,v,w), 
 $$
Then
$$
\delta(u,v,w)=\gamma(u,v,w)+\gamma(w,v,u).
$$
A polarisation $\beta(u,v)-\beta(v,u)=\omega(u,v)$ is normalised if it satisfies 
$\beta(v,0)=0$ for all $v\in V$.}

\medskip

Since $\beta(u,v)-\beta(v,u)=\omega(u,v)$ and $\omega(0,v)=\omega(v,0)=0$, 
for each polarisation we have $\beta(v,0)=\beta(0,v)$. 

\medskip

{\bf 6.5.3. Remark.} {\it  Unlike  the symplectic case recalled in the previous
subsections, in the almost--symplectic setting 
$\omega$ and $\beta$ are not multilinear, as that would imply the cocycle
condition (the vanishing of $\delta$ and $\gamma$). }

\medskip

{\bf 6.5.4. Definition.}{\it We define the following functions that measure 
lack of linearity of $\beta$ in the left/right variable:
$$ \gamma_\ell(u,v,w):= \beta(u+v,w)-\beta(u,w)-\beta(v,w) $$
$$ \gamma_r(u,v,w): =\beta(u,v+w)-\beta(u,w)-\beta(v,w), $$
so that we can write
$$ \gamma(u,v,w)=\gamma_r(u,v,w)-\gamma_\ell(u,v,w), $$
and similarly for $\delta_\ell(u,v,w)$ and $\delta_r(u,v,w)$, measuring
the lack of linearity of $\omega$.
}

\medskip

The code loops we consider here are obtained as follows.

\medskip

{\bf 6.5.5. Definition.} {\it The almost--symplectic code loops $\cL(V,\omega)$ 
and $\cL(V,\beta)$
over $\F_q$
are defined as follows.

\smallskip

(i) If $q$ is odd, such a loop is an extension
$$ 
0 \to \F_q \to \cL(V,\omega) \to V \to 0, 
$$
where $(V,\omega)$ is an almost--symplectic vector space over $\F_q$. 

\smallskip

(ii) If $q=2^r$, it is an extension
$$
0 \to R \to \cL(V,\beta) \to V \to 0, 
$$
where $(V, \omega)$ is an almost--symplectic vector space $(V,\omega)$  with
polarization $\beta$ over $\F_{2^r}$.

\smallskip

The non--associative multiplication is given, in the first case, by
$$ 
(u,x)\star (v,y)=(u+v, x+y+\frac{1}{2} \omega(u,v)), 
$$
and in the second case by
$$
 (u,x)\star (v,y)=(u+v, x+y+\beta(u,v)). 
 $$}

\medskip

The case with $q$ odd can be seen as a special case of existing construction of
loops described in [Hsu00a]  and [Hsu00b]. Thus,
we focus on the case of characteristic $2$, which is different. We start with a
characterization of the Moufang condition for the loops $\cL(V,\beta)$.

\smallskip

The following argument was suggested to us by the referee, based on the
construction of free Moufang loops in the variety generated by code loops of [GriPir18].

\medskip

We note that a Moufang loop of the form $\cL = \cL(V,\beta)$ is a particular case of the Moufang loop from the variety $\cE_4$, 
given by the set of identities
$$  x^8 = [x^2,y] = (x^2,y,z) = ([x,y],z,t) = (x,y,z)^2 =
[[x,y],z] = [x,y]^2 = 1 \, .    \eqno(6.1) $$

\smallskip

Let $\phi:  \cL(V,\beta) \to V$ be the epimorphism from Definition~6.5.5 and let
$\{ v_i'\,|\, i=1,\ldots, n \}$ be a subset of $\cL(V,\beta)$ such that $\{ v_i = \phi(v_i') \,|\, i = 1, \ldots ,n \}$
is a basis of $V$. We denote by $\cL'$ a subloop of $\cL(V, \beta)$ generated by $\{ v_i' \,|\, i = 1, \ldots ,n \}$.
We have $\cL(V,\beta)=\cL'$ iff $\roman{Ker}(\phi)=A(\cL(V,\beta))$, where $A(\cL)$ is the minimal normal
subloop of a loop $\cL$ such that $\cL/A(\cL)$ is an elementary abelian group of exponent two.
The restriction of $\phi$ to $\cL'$ is an epimorphism. In general, it is not true that $\cL(V, \beta) = \cL' \times Z_1$, 
for some subgroup of $\roman{Ker}(\phi)$. However, we can find an epimorphism $\alpha: (\Z/4\Z)^m \times \cL' \to \cL(V,\beta)$
with minimal $m$, such that the restriction of $\alpha$ to $\cL'$ is $\phi$.

Thus, we can reduce the problem of describing the loops of the form $\cL(V, \beta)$ to the problem of describing 
the free loops in the variety $\cE_4$. Let $F_n$ denote a free loop with $n$ generators in the variety $\cE_4$.

\medskip

{\bf 6.5.6. Construction of $F_n$.}
Let $\{ x_1, \ldots ,x_n \}$ be a set of free generators of the free loop $F_n$. Let
$\dim_{\F_2} V = n$. We define an abelian group $Z = \oplus_{i=1}^n (\Z/4\Z) z_i  \oplus 
V \wedge V \oplus V \wedge V \wedge V$. We can consider $V \wedge V \oplus V \wedge V \wedge V$
as an $\F_2$-vector space with a basis $\{ w_{ij}=v_i \wedge v_j\, ;\, u_{ijk}=v_i\wedge v_j \wedge v_k \, |\, 1 \leq i < j < k \leq n \}$,
for some fixed basis $\{ v_1, \ldots, v_n \}$ of $V$ over $\F_2$. 

\smallskip

We denote by $x_\sigma$ the elements $x_\sigma := (\cdots (x_{i_1} x_{i_2} )\cdots )x_{i_s} \in F_n$ and 
we take $v_\sigma :=v_{i_1} + v_{i_2} + \cdots +v_{i_s}$, for a set of indices $\sigma=\{ i_1 < \cdots < i_s \}$.
We then define a cocycle $\tau: V \times V \to Z$ by setting
$$ \tau(v_\sigma, v_\mu) := ( \prod_{i\in \sigma\cap \mu} z_i )\, ( \prod_{i>j, i\in\sigma, j\in \mu} w_{ij} )\, ( \prod_{i\in \sigma, j<k \in \mu} u_{ijk} )\, . $$

\medskip

{\bf 6.5.7. Theorem.} {\it The loop $\cL(V,\tau)$ is Moufang and is isomorphic to a free loop $F_n$ in the variety $\cE_4$.}

\medskip

{\it Proof.} First we check that $\tau$ is a Moufang cocycle. Let $N$ be a subloop of
$\cL = \cL(V,\tau)$ generated by $\{ z_1^2, \ldots , z_n^2 \}$. It is clear that $\cI \simeq \F_2^n$ 
and $\cL/N \simeq E_n$, where $E_n$ is a free Moufang loop on $n$ generators in the variety $\cE$ 
defined in [GriPir18]. Let $f(x,y,z) = ((xy)(zx))(x((yz)x))^{-1}$. Then $\cL$ is a Moufang loop iff $f(\cL,\cL,\cL) = 1$.
We showed that $f(\cL,\cL,\cL) \subseteq N$. On the other hand, if $\cI = V \wedge V \oplus V \wedge V \wedge V \subset Z \subset \cL$, 
then $\cL/\cI \simeq (\Z/8\Z)^n$ is a group. Thus, $f (\cL, \cL, \cL) \subset N \cap \cI = 1$ hence $\cL$ is a Moufang loop.
Then, since $F_n$ is a free loop in the variety $\cE_4$ and $\cE\subset \cE_4$, there exists an epimorphism
$\lambda: F_n \to E_n$ with kernel $\roman{Ker}(\lambda) \simeq \F_2^n$ generated by $x_1^4, \ldots, x_n^4$. Thus, we have
$\roman{Ker}(\lambda) \simeq N$ and $| F_n |=| \cL(V,\tau) |$. Since $F_n$ is a free loop,  
there exists an epimorphism $\mu: F_n \to \cL(V,\tau)$, so that we obtain $F_n \simeq \cL(V,\tau)$. 
$\blacksquare$

\medskip

{\bf 6.5.8. Corollary.} {\it Let $\cL(V,\beta)$ be a Moufang loop with $n$ generators obtained through the construction of Definition~6.5.5.
Then $|\cL(V,\beta)|\leq 2^t$ where $t=n(n^2 + 8)/3$.}

\smallskip

We can then obtain a characterization of loops $\cL(V,\beta)$ that satisfy the Moufang condition, using the previous results.
Indeed, Theorem~6.5.7 allows us to describe all Moufang loops of the form $\cL(V, \beta)$ for a fixed choice of a $\F_2$-vector 
space $V$ and a natural number. 

\smallskip

 We can assume that $\cL(V,\beta)\neq Q \times \cL'$, where for $\phi: \cL(V,\beta)\to V$ as above, we have $\phi(\cL')=V$ and
 $\phi(Q)=0$. In this case, we say that $\cL(V,\beta)$ is {\it irreducible}.
 
\smallskip 

We can then construct a non-splitting extension $\F_n$ of $F_n$ in the following way.  If $P=(\Z/4\Z)^m \times F_n$, with 
$m = (n-1)n(n+1)/6$, let $\{ t_{ij}; t_{ijk}\, |\, 1\leq i<j<k\leq n\}$ be a basis of the free $\Z/4\Z$-module $(\Z/4\Z)^m$. 
We denote by $J$ a normal subloop of $P$ generated by $\{ w_{ij} t_{ij}; u_{ijk} t_{ijk} \,|\, 1 \leq i < j < k \leq n \}$. By construction
we have $\F_n =P/J$. We then have the following characterization of the Moufang property for loops $\cL(V,\beta)$.

\medskip 

{\bf 6.5.9. Proposition.} {\it Let $\cL(V,\beta)$ be an irreducibe Moufang loop obtained as in Definition~6.5.5. 
Then there exists an epimorphism $\mu : \F_n \to \cL(V,\beta)$ with central kernel.} 

\medskip

{\it Proof.} Let $\phi: \cL(V,\beta)\to V$ be an epimorphism with $\{ \phi(v_i')\,|\, i=1,\ldots, n \}$ a basis of $V$. 
We denote by $\cL'$ a subloop of $\cL(V,\beta)$ generated by the set $\{ v_i' \,|\, i=1,\ldots, n \}$. Since $F_n$ 
is a free loop, there exists an epimorphism $\pi: F_n \to \cL'$. Let 
$$ W=\roman{Ker}(\phi) \cap \cL' =(\Z/4\Z)^s \times (\Z/2\Z)^t \subseteq R=(\Z/4\Z)^r \, . $$
By the structure of $F_n$ we get that $s \leq n$ and  $t \leq m = (n-1)n(n+1)/6$. 
There exists a subgroup $T$ of $\roman{Ker}(\phi)$ such that $W \subset T = (\Z/4\Z)^{s + t}$ and 
an epimorphism $\mu : \F_n \to \cL' \cdot W$. If $\cL' \cdot W\neq \cL(V,\beta)$, then the loop $\cL(V,\beta)$
would be reducible. 
$\blacksquare$

\medskip

{\bf 6.6. Quantum codes from almost--symplectic code loops.} We come now to extending the
result of Proposition~6.2.3 to our construction of loops $\cL(V,\beta)$. To this purpose, we first need
to recall the appropriate notion of linear representations of loops, then we need to introduce 
isotropic subspaces, and then obtain from them the respective CRSS quantum codes.

\medskip

{\bf 6.6.1. Linear representations of loops.} 
A notion of linear representations of loops was developed in [Log93]. It is closely related to
the Eilenberg notion of representation for non--associative algebras [Eil48].

\smallskip

Given a loop $\cL$ and a vector space $\cH$ over a field $F$, left and right composition
maps are defined as $\ell,\rho: \cL \to Aut(\cH)$, which we write simply as
$$ 
\ell_a(h)=a\star h, \ \ \  \rho_a(h)=h \star a. 
$$
These maps should satisfy $a\star (h+h')=a\star h + a\star h'$, $(h+h')\star a=h\star a + h'\star a$,
$a\star (\lambda h) =\lambda \, a\star h$, $(\lambda h)\star a=\lambda\, h\star a$, for all $a\in \cL$,
$h,h'\in \cH$, $\lambda\in F$. One defines on $\cL\times \cH$ the multiplication
$$ 
(a,h)\star (b,h')=(a\star b, a\star h' + h \star b). 
$$
We also define the associator $$[a,b,h]=(a\star b) \star h-a\star (b\star h)$$ 
for $a,b\in \cL$ and $h \in \cH$.

\smallskip

Over a field $F$, one can associate to a loop $\cL$ the non--associative algebra $F[\cL]$,
the analog of the associative group algebra for groups. The maps $\ell,\rho$ of a representation of $\cL$ on
an $F$--vector space $\cH$ extend by linearity to $F[\cL]$, in the sense of representations of
non-associative algebras, [Eil48]. 

\smallskip

If the loop $\cL$ satisfies the Moufang identity, then the maps $\ell,\rho: \cL \to Aut(\cH)$ of
a representation of $\cL$ must satisfy the following conditions (see [Log93]):
the associator $[a,b,h]$ is skew-symmetric for all $a,b\in F[\cL]$ and $h\in \cH$;
the identities $h\star (b\star(a\star b))=((h\star b)\star a)\star b$ and $((a\star b)\star a) \star h =a\star (b\star (a\star h))$
hold, for all $a,b\in F[\cL]$ and
all $h\in \cL$. 

\medskip

{\bf 6.6.2. Isotropic subspaces.} 
As above, denote by $V$ be a vector space over $\F_{2^r}$, put $R=\cO_K/\fm_K^2$ where $\cO_K/\fm_K=\F_{2^r}$,
and assume that $V$ is endowed with an almost--symplectic structure $\omega: V\times V\to R$ with
normalised polarisation $\beta$.

\medskip

{\bf 6.6.2.1. Definition.} {\it 
An isotropic subspace $C\subset V$ is a linear subspace where the almost symplectic form 
vanishes identically, $\omega|_C = 0$. A polarisable subspace 
$P\subset V$ is a linear subspace for which there is an
enhancement function $\alpha: P \to R$ satisfying
$$ 
\alpha(u+v)-\alpha(u)-\alpha(v)=\beta(u,v), \ \ \  \forall u,v \in P. 
$$
A polarized subspace is a pair $(P,\alpha)$ satisfying the condition above.}

\medskip

The polarization relation is just the Hochschild coboundary relation $\beta=d\alpha$, hence
it implies $\gamma|_P =d\beta|_P =0$. 

\medskip

{\bf 6.6.2.2. Proposition.} {\it A polarized subspace $(P,\alpha)$ determines a section $\tau: P \to \cL(V,\beta)$
of the projection $\cL(V,\beta)\to V$, with image $\tau(P)\subset \cL(V,\beta)$ a subgroup of the loop $\cL(V,\beta)$. 
If $P$ is also isotropic, then $\tau(P)\subset \cL(V,\beta)$ is an abelian subgroup. }

\medskip

{\it Proof.} The section $\tau: P \to \cL(V,\beta)$ is constructed as in the symplectic case of [GuHa12], by
taking $\tau(v)=(v,\alpha(v))$ for $v\in P$. This satisfies
$$ (v,\alpha(v))\star (w,\alpha(w))=(v+w,\alpha(v)+\alpha(w)+\beta(v,w))=(v+w,\alpha(v+w)). $$
This multiplication is associative since $d\beta|_P=0$.
On an isotropic subspace the polarization $\beta$ is symmetric, hence the resulting multiplication is also commutative. 
$\blacksquare$

\medskip

{\bf 6.6.3. CRSS quantum codes from almost--symplectic loops.} We consider here  almost--symplectic
loops $\cL(V,\omega)$ in characteristic $p>2$ and $\cL(V,\beta)$ in characteristic $p=2$, as in
Definition~6.5.5. We simply write $\cL$ for the loop when both cases are considered. 

\smallskip

Let $\cH=\C[\cL]$ be the complex vector space of complex valued functions on $\cL$, endowed with
 the left and right composition maps $\ell,\rho: \cL \to Aut(\cH)$, as in Section~6.6.1, given by
the left and right action of $\cL$ on itself extended by linearity. 
We write $| a \rangle$ with $a\in \cL$ for the canonical basis of $\cH$.

\smallskip

Given a character $\chi: Z(\cL)\to \C^*$ (that is, a character $\chi: R \to \C^*$ for $p=2$ or $\chi: \F_q\to \C^*$ for $p>2$),
let $\cH_\chi\subset \cH$ be the subspace of functions $f: \cL \to \C$ that transform like
$\ell_{(0,x)} f (u,y)= \chi(x) f(u,y)$, for $x\in Z(\cL)$ and $(u,y)\in \cL$. 

\smallskip

When $p=2$, a polarized isotropic subspace $(C,\alpha)$ is a pair of an
isotropic subspace $C \subset V$ together with an enhancement function as in Definition~6.6.2.1 above.
When $p>2$ let $C\subset V$ be an isotropic subspace. In the following for simplicity we will refer to 
both cases simply as ``an isotropic subspace", with the function $\alpha$ implicitly understood in 
the characteristic $2$ case.

\medskip

{\bf 6.6.3.1. Theorem.}
{\it An isotropic subspace $C \subset V$ determines a commuting family of error operators
$\chi(\tau(v)) E_v$, with $v\in C$, and an associated error correcting quantum code 
$\cC_C\subset \cH_\chi$ given by a joint eigenspace of these operators. The assignment
$C\mapsto \cC_C$ is the almost--symplectic CRSS algorithm. }

\medskip

{\it Proof.} The left composition map $\ell: \cL \to Aut(\cH_\chi)$ induces a representation $\pi: \tau(C) \to Aut(\cH_\chi)$ of
the abelian subgroup $\tau(C)\subset \cL$, as in Proposition~6.6.2.2. We can write the operators on $\cH_\chi$ obtained
in this way as $\pi(v,\tau(v))=\chi(\tau(v)) E_v$, and regard them as a commuting family of error operators on $\cH_\chi$.
A common eigenspace of the $\chi(\tau(v)) E_v$ in $\cH_\chi$ gives a subspace $\cC_C \subset \cH_\chi$ that is
the CRSS quantum code associated to the classical code $C\subset V$ through the code loop $\cL$. 
$\blacksquare$

\medskip

{\bf 6.7. Locally conformally symplectic structures and perfect tensors.}
We now discuss how to generalize Proposition~6.3.2 to the case of
code loops. 

\smallskip

In general, in the almost--symplectic case, the fact that $\delta=d\omega\neq 0$ means that we
do not have a Darboux decomposition of $(V,\omega)$, hence $\omega$ by itself does not
determine an explicit identification of of $\cH_\chi$ with a tensor product of $q$--ary qubits.  
Thus, in the almost--symplectic setting one needs to consider special
cases, such as an analog of the conformally flat almost--symplectic structures on manifolds, from
which a decomposition of the space into $2$--dimensional Darboux pieces can still be obtained. 

\smallskip

When a decomposition into a product of qubits is given, one can again use as in [HMPS18] Lagrangians
in general position with respect to this decomposition (enhanced Lagrangians $(L,\alpha)$ in the case
of characteristic $2$) to obtain perfect tensors through the same kind of CRSS construction described above. 

\smallskip

We focus here in particular on a
case modelled on manifolds with locally conformally symplectic structures,
for which a Darboux theorem holds, see [OtiSta17]. 

\medskip

{\bf 6.7.1. Definition.} {\it Let $V$ be a finite dimensional vector space over $\F_q$.

\smallskip

 An $1$--form is given by a function\,\, $\theta : V\to A$, and a $2$--form is given
 by a function $\omega : V\times V \to A$, where $A = \F_q$, if $q$ is odd, and $A = R$,
 if $q$ is even.
\smallskip

Define the wedge product  $\theta \wedge \omega$ as the function of three arguments
$$
 (\theta \wedge \omega)\, (u,v,w): = \theta(u)\, \omega(v,w) + \theta(w)\, \omega(u,v). 
$$
}

\medskip

This definition is compatible with defining the wedge product of two $1$--forms $\theta_1,\theta_2$ as 
$$ (\theta_1\wedge \theta_2)(v,w):=\theta_1(v)\theta_2(w)-\theta_1(w)\theta_2(v), $$
through the expected relation
$$
 d(\theta_1\wedge \theta_2) =d\theta_1 \wedge \theta_2 -\theta_1 \wedge d\theta_2 . 
$$

\medskip

{\bf 6.7.2. Definition.} {\it 
Let $V$ be a vector space over $\F_q$. An almost--symplectic form $\omega$ on $V$
is called  a {\it locally conformally symplectic structure} if there is a closed $1$--form $\theta$
such that
$$
 d\omega = \theta \wedge \omega . 
$$
Moreover, $\theta$ and $\omega$ must be homogeneous with respect
to scalar multiplication on $V$.}

\medskip

Consider an almost symplectic vector space $(V,\omega)$ over $\F_q$, and the
associated loop $\cL$ (that is, $\cL(V,\omega)$ for characteristic $p>2$ and
$\cL(V,\beta)$ in characteristic $p=2$). Let $L\subset V$ be a Lagrangian with
respect to $\omega$ (an enhanced Lagrangian $(L,\alpha)$ for $p=2$) and 
let $\tau(L)\subset \cL$ be the resulting subloop, with $\tau(L)=\{ (v,0)\,|\, v\in L\}$
for $p>2$ and $\tau(L)=\{ (v,\alpha(v))\,|\, v\in L\}$ for $p=2$. 
By Proposition~6.6.2.2 we know that $\tau(L)$ is in fact an abelian subgroup. 

\smallskip

Let $\cH(V,L,\omega)\subset \C[\cL]$
be the subspace of functions $f(u,x)$ that are invariant under the action of $\ell(\tau(L))$, through the
left composition map $\ell$ of the loop representation. Let $\cH_\chi(V,L,\omega)$ be the
subset of functions that also transform as $\ell_{(0,y)} f(u,x)=\chi(y) f(u,x)$, under a character
$\chi: Z(\cL\to \C^*$ (that is, $\chi: \F_q \to \C^*$ for $p>2$ and $\chi: R\to \C^*$ for $p=2$). 

\medskip

{\bf 6.7.3. Proposition.} {\it A locally conformally symplectic structure $(V,\omega)$
over $\F_q$ determines a decomposition into qubits, 
$\cH\simeq \otimes_i \cH_i$ with $\cH_i\simeq \C^q$, of $\cH=\cH_\chi(V,L,\omega)$.}

\medskip

{\it Proof.}
Since $\theta$ is homogeneous, the closedness $d\theta=0$ means that
$d\theta (u,v)=\theta(v)-\theta(u+v) +\theta(u) =0$,
that is, $\theta$ is linear. Thus, we can decompose the vector space $V$ into
the kernel $K=Ker(\theta)$ and an one--dimensional complement, $V=K\oplus \F_q$,
satisfying the condition $d\omega|_K \equiv 0$. 

\smallskip

Since $\omega$ is non--degenerate,
one can find a pair of vectors $u,v$ in $K$
such that $\omega(u,v)\neq 0$,  Since $\omega$ is
closed on $K$, one can then decompose $K$ into this two-dimensional subspace and
a complement $W=\{ w\in K\,|\, \omega(u,w)=\omega(v,w)=0 \}$. One can proceed in the
same way by restricting $\omega$ to $W$, and obtain in this way a decomposition of
$K$ into subspaces $K_i\simeq \F_q^2$, with $K\simeq \oplus_i K_i \oplus \F_q$. 
This provides an overall decomposition of $V\simeq \oplus_i V_i$ with $V_i\simeq \F_q^2$.
The direct sum $V=\oplus_i V_i$ with $\omega_i=\omega|_{V_i}$ gives a corresponding
decomposition of the complex vector space $\cH=\otimes_i \cH_i$ with each $\cH_i\simeq \C^q$
a single qubit space. 
$\blacksquare$

\medskip

We refer to the decomposition of the locally conformally symplectic space $(V,\omega)$ obtained in this
way and the corresponding decomposition of $\cH$ into qubits as the {\it Darboux decomposition.}

\smallskip

We thus obtain the generalisation of Proposition~6.3.2 to the case of
code loops, by the same argument as in [HMPS18].

\medskip

{\bf 6.7.4. Theorem. } {\it A Lagrangian $L$ that is in general position
with respect to the Darboux decomposition of the locally conformally symplectic structure,
determines a perfect tensor in $\cH$. }

\medskip

{\bf 6.8. Networks of perfect tensors.} We now show that the construction of
perfect tensors associated to almost--symplectic code loops with a
locally conformally symplectic structure and a Lagrangian in general position,
as in Theorem~6.7.4 can be used to construct networks of perfect tensors
associated to certain combinatorial structures that arise from the relation between
Moufang loops and Latin square designs.

\smallskip

In particular, we use this construction
of networks of perfect tensors to show that the Latin square designs obtained from
our code loops have an associated information--theoretic entropy functional. 

\smallskip

We will first introduce tensor networks and the associated entanglement entropy.
We will then review the relation between Moufang loops and Latin square designs
and present our construction of networks of perfect tensors. We then conclude the
section with some questions on the construction of tensor networks on chamber
systems and on their universal $2$--covers, when the latter are buildings.

\medskip

{\bf 6.8.1. Tensor networks and entanglement entropy.} 
A tensor network is a pattern of contraction of indices of tensors. This can be
stated more precisely as follows.

\smallskip

We will encode combinatorics of  finite graphs by identifying each such graph $G$
with a quadruple $G=(F,V,\partial,j)$, where $F$ is the set of flags (half--edges), $V$ the
set of vertices, $\partial$ the boundary map $\partial: F \to V$ that identifies the root vertex of
each flag, and $j$ is the structure involution $j: F \to F$, $j^2=id$, that describes how half--edges are glued
together into edges of $G$. Using the physics terminology, we call {\it internal edges} those
pairs $e=(f,f')$ with $f\neq f'$ and $f'=j(f)$, and {\it external edges} the flags $f$ that are fixed
by the involution: $j(f)=f$. 

\smallskip

Much more details can be found in [BoMa07], Sec. 1, in particular, a description of
morphisms of graphs and other information, which we will use below
without repeating the definitions.

\medskip

{\bf 6.8.1.1. Definition.} {\it 
A tensor network $(G,\cH,T)$ consists of a finite graph $G$ as above, without multiple edges, where the vertices $v\in V$ 
are decorated by pairs $(\cH_v,T^{(v)})$ of a complex vector space $\cH_v= (\C^q)^{\otimes \deg(v)}$, 
for some $q=p^r>0$ a power of some prime $p$, with $\deg(v)$ the valence of the vertex, and a $T^{(v)}\in \cH_v$.

\smallskip

We can view such $T^{(v)}$ as a tensor $T^{(v)}=(T^{(v)})_{i_1,\ldots, i_{\deg(v)}}$,
with indices  $i_f \in \F_q$, labelled by the flags $f\in F$ with $\partial(f)=v$.
An edge $e=(f,f')$, $f'=j(f)$, 
with $\partial e=\{ v, v' \}$ corresponds to a contraction of indices of the tensors $T^{(v)}$ and $T^{(v')}$ of the form
$$ \sum_{i_f, i'_{f'}\in \F_q} \delta^{i_f, i'_{f'}}\, T^{(v)}_{i_1,\ldots, i_{\deg(v)}} \, 
T^{(v')}_{i'_1, \ldots, i'_{\deg(v')}}, $$
with $\delta^{ij}$ the Kronecker delta function. The internal edges of $G$ are called the {\it bonds}
of the tensor networks. The external edges of the graph $G$ correspond to indices of the
tensors that remain non--contracted. We call them the {\it dangling legs} of the tensor networks.
The graph $G$ is called the {\it support} of the tensor network. 
}

\medskip 

{\bf 6.8.1.2. Definition.} {\it Let $G$ be a finite connected graph.
A cut--set of $G$ is such a subset $C\subset E_{in}(G)$ of the set of 
internal edges, that if all the edges $e\in C$ are cut, the graph
$G$ is split into exactly two non--empty 
connected components, $G\smallsetminus C=G_{C,1}\sqcup G_{C,2}$.
}

\medskip

{\bf 6.8.1.3. Lemma.} {\it Let $G$ be a finite connected graph and let $E_{in}(G)$ and $E_{ext}(G)$ be the sets of
internal and external edges of $G$.
A tensor network $\cT=(G,\cH,T)$ computes an entangled state $|\psi_\cT\rangle$ 
in the space $\cH_\cT=(\C^q)^{\otimes |E_{ext}(G)|}$,
with $|E_{ext}(G)|$ the number of external edges of the graph $G$. 
In the case where $E_{ext}(G)=\emptyset$, this computation just gives a complex number, 
the amplitude $\alpha_\cT$. Given a cut-set $C$, one obtains entangled states $|\psi_{C,i}\rangle$ 
in $(\C^q)^{\otimes |C|}$, 
associated to the restrictions of the tensor network to the components $G_{C,i}$, satisfying
$|\alpha_\cT|=|\langle \psi_{C,1}, \psi_{C,2} \rangle|$.
}

\medskip

{\it Proof.} Consider
the standard basis $| a_1\ldots a_N \rangle$ of the space $(\C^q)^{\otimes N}$ of $N$ $q$-ary qubits,
with $a=(a_1,\ldots, a_N)\in \F_q^N$ and $a_i\in \F_q$, and $| a_1\ldots a_N \rangle=| a_1\rangle \otimes 
\cdots\otimes| a_N \rangle$. At each vertex $v\in V(G)$ we obtain an entangled state
$$ | \psi_v \rangle =\sum_{a_1,\ldots, a_{\deg(v)} \in \F_q} T^{(v)}_{a_1, \ldots, a_{\deg(v)}} | a_1\ldots a_{\deg(v)} \rangle, $$
obtained as a superposition of the pure states $| a_1\rangle \otimes 
\cdots\otimes| a_{\deg(v)} \rangle$. Contracting two tensors $T^{(v)}$ and $T^{(v')}$ along an
edge $e$ with $\partial(e)=\{ v, v'\}$ gives rise to an entangled state that is a superposition of the
pure states associated to the remaining dangling legs at the two vertices,
$$ |\psi_e \rangle=\sum_{a_i, b_j\in \F_q} \delta^{a_f, b_{f'}}\, T^{(v)}_{a_1,\ldots, a_{\deg(v)}} \, 
T^{(v')}_{b_1, \ldots, b_{\deg(v')}} | \hat a^{(f)}, \hat b^{(f')}\rangle, $$
where $\hat a^{(f)}=(a_1,\ldots, \hat a_f, \ldots, a_{\deg(v)})$ and $\hat b^{(f')}=(b_1,\ldots,
 \hat b_{f'}, \ldots, b_{\deg(v')})$, 
and $\hat a_f$ and $\hat b_{f'}$ means that this entry in the vector has been removed.
In a similar way, performing the contractions of the tensor indices 
along the edges of the graph $G$ gives rise to an entangled state $| \psi_G \rangle$
that is a superposition of the pure states associated to the dangling legs of $G$
$$ |\psi_G \rangle=\sum_{c_1,\ldots, c_N \in \F_q} \tau_{c_1,\ldots, c_N} |c_1\ldots,c_N\rangle, $$
where $N=\# E_{ext}(G)$ is the number of external edges. The coefficients $\tau_{c_1,\ldots, c_N}$
are computed by performing all the contraction of indices across all the internal edges of the graph $G$. 

If $G$ has no external edges, each edge $e\in C$, seen as a pair $e=(f_1,f_2)$ with
$f_2=j(f_1)$ and $\partial(f_i)\in G_{C,i}$, endows both  components $G_{C,i}$ with
an external edge, so that the total number of such edges is 
$| E_{ext}(G_{C,i}) |=|C|$, for both $i=1,2$.  One can then consider the states $|\psi_{C,i} \rangle$
computed by the tensor network as above. The amplitude $\alpha_\cT$ is obtained from
these by contracting the indices corresponding to the pairs $(f_1,f_2)$. 
$\blacksquare$

\medskip

A tensor network $\cT=(G,\cH,T)$ with the associated
entangled state $|\psi_\cT\rangle$ in $\cH_\cT=(\C^q)^{\otimes |E_{ext}(G)|}$ as
above determines a corresponding density matrix, written in bra-ket notation as
$$ \rho =\frac{1}{\langle \psi_\cT | \psi_\cT \rangle} \,\, |\psi_\cT\rangle \, \langle \psi_\cT |. $$
Given a partition $A\sqcup B$ of the set of external edges of $G$, we can consider
$$ \rho_A =\Tr_B (\rho) $$
with $\Tr_B: \cH_A\otimes \cH_B \to \cH_A$, so that $\rho_A$ is obtained from $\rho$ 
by tracing out (contracting the indices of) the dangling legs in $B$. 

\medskip

{\bf 6.8.1.4. Definition.} {\it The entanglement
entropy of the tensor network $\cT=(G,\cH,T)$ is then given by the assignment
$$ A \mapsto S_\cT(A):= \Tr (\rho_A \log \rho_A), $$
for $A\subset E_{ext}(G)$ ranging over all subsets of external edges.  

In the case of a connected graph $G$ with no external edges, the
entanglement entropy of the tensor network $\cT=(G,\cH,T)$ is
given by the assignment 
$$ A_i \mapsto S_{\cT,C,i} (A_i):= \Tr (\rho_{C,A_i}\log \rho_{C,A_i}), $$
for $C$ ranging over cut--sets and $A_i \subset E_{ext}(G_{C,i})$ ranging over all 
subsets of external edges of the components $G_{C,i}$, and with
$\rho_{C,A_i}=\Tr_{C\smallsetminus A_i} (\rho_{C,i})$ where
$\rho_{C,i}$ is the density matrix associated to the entangled state $|\psi_{C,i}\rangle$.
}

\medskip

{\bf 6.8.2. Moufang loops and Latin square designs,}
We recall here briefly some notions from combinatorial designs and the geometry of buildings,
closely related to loops. We refer the reader to [Cam03], [Hall19], [MeiStWe13] 
for more details. 

\medskip

{\bf 6.8.2.1. Definition.} {\it A {\it Latin square design} is a pair $\cD=(P,A)$. 

\smallskip

Here $P$
is a set of $3N$ points, represented as a disjoint union 
$P=P_1\sqcup P_2\sqcup P_3$ of three subsets of cardinality $N$.

\smallskip

$A$ is a family of subsets of $P$, called lines, 
with the property that each line in $A$ contains exactly $3$ points, one from each
of the three subsets  $P_i$, and such that any two points from two different subsets  $P_i$
belong to exactly one line in $A$. 

\smallskip

The Latin square of the design $\cD$ is the
$N\times N$ -- matrix with entries corresponding to the $N^2$ lines in $A$ and with
$(x_1,x_2)$--entry equal to $x_3$ if the line containing $x_1\in P_1$ and $x_2\in P_2$ has $x_3\in P_3$
as the third point. The order of a Latin square is the number $N$ of points of each type.}

\medskip

Latin square designs form a category with objects $\cD=(P,A)$ and morphisms $\cD\to \cD'$ given by
a triple of maps $\alpha_i: P_i \to P'_i$ such that,
if $(x_1,x_2,x_3)$ is a line in $A$ then $(\alpha_1(x_1),\alpha_2(x_2),\alpha_3(x_3))$ is a line in $A'$.

\smallskip

Given a loop $\cL$, the {\it Thomsen design} $\cD(\cL)$ has set of points $P=\cL_1\sqcup \cL_2 \sqcup \cL_3$,
three copies of $\cL$ labelled $i=1,2,3$, and set of lines $A=\{ (x_1,x_2,x_3)\,|\, (x_1\star x_2)\star x_3 =1 \in \cL \}$. Conversely, given any Latin square design $\cD$, there is a loop $\cL(\cD)$ with this property, 
the {\it Thomsen loop} of $\cD$.
The Thomsen loop assignment $\cD \mapsto \cL(\cD)$ is functorial and gives an equivalence of categories
between the category of Latin square designs and the category of loops, where objects are loops $\cL$
and morphisms are isotopisms, namely triples of maps $(\alpha,\beta,\gamma): \cL \to \cL'$ satisfying 
$\alpha(x)\star' \beta(y)=\gamma(x\star y)$ for all $x,y\in \cL$, see Theorem~3.4 of [Hall19]. 

\smallskip

An {\it automorphism} of a Latin square design $\cD=(P,A)$ is a permutation of $P$ that sends lines to lines. A central
automorphism $\tau_x$ of $\cD$, {\it centered} at a point $x\in P$ is an automorphism that fixes $x$ and
exchanges the remaining two points on each line in $A$ containing $x$  (see Section~3.2 of [Hall19]).

\medskip

{\bf 6.8.2.2. Definition.} {\it
A  central Latin square design
is a design that admits a central automorphism at every point $x\in P$.}

\medskip

The Thomsen functor restricted to this subcategory gives an equivalence between the category of 
central Latin square designs and the category of {\it Moufang loops} (Theorem~3.11 of [Hall19]).

\smallskip

A subdesign $\cD'=(P',A')$ of a Latin square design $\cD=(P,A)$ consists of sets $P'\subseteq P$ 
and $A'\subseteq A$ of points and lines that form a Latin square design.

\smallskip

Any non--empty set of lines in $\cD$ is contained in a unique minimal subdesign. 
This is referred to as the subdesign generated by the given set of lines.  

\medskip

{\bf 6.8.2.3. Lemma.} {\it Consider the almost--symplectic code loops $\cL(V,\omega)$, if characteristic $p$ is odd,
or $\cL(V,\beta)$ if $p=2$, as in Definition~6.5.5. 

\smallskip

The Thomsen design $\cD(\cL(V,\omega))$, resp.
$\cD(\cL(V,\beta))$ has an associated graph $G=G_{\cL(V,\omega)}$, resp.
$G=G_{\cL(V,\beta)}$, describing how points of the design are connected by lines, with
$\roman{card}\, V(G)=3N$ and $\roman{card}\, E(G)=3N^2$, where $N=q^{2n+1}$ for $q=p^r$ with $p$ odd and
$N=q^{2n+2}$ for $q=2^r$. 

\smallskip

The choice of an isotropic subspace $C\subset V$ with $\dim_{\F_q} C=k$ 
determines a subdesign $\cD(\tau(C))$ and a subgraph $G_{\tau(C)}$ with $3 q^k$ vertices and
$3 q^{2k}$ edges. For $p=2$ any pair of intersecting lines in $\cD(\tau(C))$ generate a subdesign of order $2$. }

\medskip

{\it Proof.} 
We can identify as sets $\cL(V,\omega)\simeq V\times \F_q$ and $\cL(V,\beta)\simeq V \times R$,
hence $\roman{card}\, \cL(V,\omega)=q^{2n+1}$, where $2n=\dim_{\F_q} V$, and $\roman{card}\, \cL(V,\beta)=2^{2nr+2r}$, where
$\dim_{\F_{2^r}} V =2n$. The Thomsen design $\cD(\cL(V,\omega))$ has $P$ consisting of three copies
of $\cL(V,\omega)$, marked with labels $i=1,2,3$, and set of lines 
$$ 
A=\{ ((u,x)_1,(v,y)_2,(w,z)_3)\,|\,\, u+v+w=0, \,\, x+y+z+\frac{1}{2}\omega(u,v) +\frac{1}{2}\omega(u+v,w)=0 \}. 
$$
The characteristic $2$ case is similar: the
Thomsen design $\cD(\cL(V,\beta)$ has $P$ consisting of three labelled copies of $\cL(V,\beta)$ and set
of lines
$$ 
A=\{ ((u,x)_1,(v,y)_2,(w,z)_3)\,|\,\, u+v+w=0, \,\, x+y+z+\beta(u,v) +\beta(u+v,w)=0 \}. 
$$

The order of the corresponding Latin square is $N=q^{2n+2}$ in the characteristic $2$ case,
with $q=2^r$ and $N=q^{2n+1}$ in characteristic $p>2$. Given a point $(u,x)_i$ in $P$, the
panel $\Pi_{(u,x)_i}$ of lines through the point $(u,x)_i$ contains $N$ lines, each containing two
other points. Two panels
$\Pi_{(u,x)_i}$ and $\Pi_{(v,y)_j}$ with types $i\neq j$ intersect in a single line. Thus we can
form a graph $G$ with set of vertices $V=P$ of uniform valence $2N$ and a single edge
between any two points with types $i\neq j$. The number of edges is $\roman{card}\, E=3N^2$.
The construction for the subgraph $G_{\tau(C)}$ is analogous. 

\smallskip

In the case of characteristic $2$, the subspace $C$, seen as an abelian group is 
an elementary abelian $2$--group and so is its image $\tau(C)\subset \cD(\cL(V,\beta))$.
As shown in Lemma~4.3 of [MeiStWe13],
the condition that any pair of intersecting lines generate a subdesign of order $2$ is equivalent
to the property that the associated loop is an elementary abelian $2$--group, hence the
property holds in this case.
$\blacksquare$

\medskip

{\bf 6.9. Networks of perfect tensors.} 
Consider the almost-symplectic code loops $\cL(V,\omega)$, in characteristic $p$ odd,
or $\cL(V,\beta)$ in characteristic $p=2$, as in Definition~6.5.5.
We assume in both cases that the almost-symplectic form $\omega$ is a locally
conformally symplectic structure as in Definition~6.7.2. Let $L$ (respectively, $(L,\alpha)$)
be a Lagrangian (respectively, enhanced Lagrangian) that is in general position with
respect to the Darboux decomposition of the conformally symplectic structure, and let
$T_L$ be the associated perfect tensor, as in Theorem~6.7.4.

\smallskip

We now construct a tensor network associated to the design $\cD(\cL(V,\omega))$
or $\cD(\cL(V,\beta))$ and its subdesign $\cD(\tau(L))$, for the chosen Lagrangian.

\medskip

{\bf 6.9.1. Proposition.} {\it Let $(V,\omega)$ be an almost--symplectic vector
space with $\dim_{\F_q} V=2n$ and $\omega$ locally conformally symplectic.
The choice a Lagrangian $L$ in general position with
respect to the Darboux decomposition gives rise to a network of perfect tensors 
$(G,\cH,T)$ with support $G\subset G_{\tau(L)}$ a uniform subgraph with
$V(G)=V(G_{\tau(L)})$ and valence $2n$, and with $\cH=\cH(V)$ with
$T\in \cH$ the perfect tensor $T=T_L$.}

\medskip

{\it Proof.} We write here $\cL$ for either $\cL(V,\omega)$ or $\cL(V,\beta)$ in odd/even
characteristic. Consider as in Lemma~6.8.2.3 the graph $G_\cL$ with the subgraph
$G_{\tau(L)}$. As support of the tensor network we consider a subgraph $G\subset
G_{\tau(L)}$ with the same set of $q^n$ vertices and with the set of edges obtained
as follows. Each vertex in $G_{\tau(L)}$ has valence $\roman{card}\, \tau(L)=2 q^n$, with the corolla 
of the vertex identified with the line segments connecting a point $u_i=(u,\alpha(u))_i$ 
of $\cD(\tau(L))$  to the remaining two points on each line in the panel $\Pi_{u_i}$. 

\smallskip

Consider now the set of lines in $\Pi_{u_i}$ that contain the points $v_j$ with $i\neq j$ with
$v=u+e_r$, $r=1,\ldots, n$ where $\{ e_r \}_{r=1,\ldots, n}$ is the standard basis of vectors in
$\F_q^n\simeq L$ with $1$ in the $r$--th entry and $0$ elsewhere. Consider as set of edges
$E(G)$ the corresponding edges of $E(G_{\tau(L)})$ connecting the points $u_i$
and $(u+e_r)_j$, for $\{ e_r \}_{r=1,\ldots, n}$. Each vertex in $G$ has valence $2 n$. 
Let $T$ be a perfect tensor in $\cH=(\C^q)^{\otimes 2n}$. We write $T=T_{\ell_1,\ldots, \ell_{2n}}$
with indices labelled by vectors $\ell=(\ell_1,\ldots, \ell_{2n})\in \F_q^{2n}\simeq V$, in the
Darboux basis, so that we have a given splitting of this set of indices into two subsets
$\ell=(\ell',\ell'')=(\ell_1,\ldots, \ell_n, \ell'_1,\ldots, \ell'_n)$.

\smallskip

We assign to each vertex $u_i$ of $G$ a copy of the space $\cH$ with the tensor $T$, so
that the indices of $T$ correspond to the $2n$ legs of the corolla of $u_i$ where we identify
the two subsets of indices with the legs connecting $u_i$ to $(u+e_r)_j$ and to $(u+e_r)_k$,
respectively, with $j,k$ the two remaining types with $(i,j,k)$ a cyclic permutation of $(1,2,3)$.
This fixes an identification of the indices of the tensor with the set of half edges at each vertex.
Each edge then corresponds to a contraction of the corresponding indices of the copies
of the tensor at the adjacent vertices.
$\blacksquare$

\smallskip

Note that, while the subgraph $G\subset G_{\tau(L)}$ has exponentially
lower connectivity (valence $2n$ rather than $2 q^n$) than $G_{\tau(L)}$,
we can still interpret 
the perfect tensor as encoding the rest of the geometry of $G_{\tau(L)}$,
through the contribution of the $T_\ell =T_{\ell_1,\ldots, \ell_n, \ell'_1,\ldots, \ell'_n}$
to the entangled state associated to the corolla of a vertex $u_i$
in $G$, given by $$ |\psi_{u_i}\rangle =\sum_{\ell} T_\ell \, |\ell \rangle, $$
where $|\ell \rangle$ is the standard basis of $\cH$ as in Definition~6.1.1.

\medskip

{\bf 6.10. Tensor networks on chamber systems and buildings.} 
In addition to the Latin square designs associated to loops,
discussed in the previous subsections, there are other
related combinatorial structures.

\medskip

{\bf 6.10.1. Definition.} {\it A {\it chamber system of type $I$} on a set $\Omega$
is a family $\{ \rho_i \}_{i\in I}$ of equivalence relations on $\Omega$ satisfying
the following conditions:

\smallskip

(i) if $\omega\sim_i \omega'$ and $\omega\sim_j \omega'$, for some $i\neq j\in I$,
then $\omega=\omega'$; 

\smallskip

(ii) the $I$-graph with vertex set $\Omega$ 
and edges $e_{\omega,\omega'}$, for $\omega\sim_i \omega'$ for some $\rho_i$, is connected.

\smallskip

Given a subset $J\subset I$
a {\it residue of type $J$} is a connected component of the $J$--graph.

The number of colors $\roman{card}\, I$ is the {\it rank} of the chamber system.}

\medskip

A graph $\Delta=(\cV,\cE,\phi)$ with vertex set $\cV$, edge set $\cE$ and an assignment of
edge colors $\phi: \cE\to I$ is a chamber system if the monochromatic subgraphs $\Delta_i$ 
with vertex set $V$ and edge set $\phi^{-1}(i)$ are a disjoint union of complete subgraphs with 
at least two vertices each (see Section~15.5 of [Hall19]). The set $V=\Omega$ is the set
of {\it chambers}, the connected components of the monochromatic subgraphs are the {\it panels}
of the chamber system. {\it Galleries} are paths in $\Delta$.

\smallskip

A {\it Latin chamber system} is a chamber system of rank $3$ where any two panels of 
different colors intersect in a unique chamber. 

\smallskip

A Latin square design determines a {\it Latin chamber system}. This has $\Omega$ given by the
set of the $N^2$ cells of the Latin square (labelled $(a,b)$ with $a,b=1,\ldots,N$), with three equivalence relations: 
(1) $(a,b)\sim_{\rho_1} (a',b')$ if $a=a'$ (same row); (2) $(a,b)\sim_{\rho_2} (a',b')$ if $b=b'$ (same column);
(3) $(a,b)\sim_{\rho_3} (a',b')$ if these cells contain the same symbol. There is only one rank $2$ residue of each type $J\subset I$ with $\roman{card}\, J=2$, which consists of an $N\times N$ grid containing all the cells. 
The set of chambers of a Latin chamber system is the set of lines of the corresponding Latin square design. 
The set of panels is the set of points of the Latin square design,
as a panel is given by the set of lines that contain a given point. 

\smallskip

A chamber system is simply $2$--connected if it is connected and each closed 
path (gallery) is $2$--homotopic to the trivial one.
The latter condition means that any closed path can be 
reduced to the trivial path through a sequence of 
replacements of subgalleries lying in rank $2$ residues by other galleries within 
the same residue. In particular, buildings are simply $2$--connected. 
Given a collection $\cC$ of closed walks in a graph $\Delta$, a $\cC$--covering $\tilde\Delta \to \Delta$
is a covering such that every closed walk in $\cC$ lifts to a closed walk in $\tilde\Delta$.
A universal $\cC$--cover exists (see Section~I.1.2.3 of [Shult11]).
A $2$-covering of a chamber system is a $\cC$-covering 
of the edge-labelled graph $\Delta$ of the chamber system
with respect to the  collection $\cC$ of all closed walks (closed galleries)
in rank $2$ residues (see Chapter~10 of [Shult11]). 

\smallskip

As shown in Proposition~4.2 of [MeiStWe13], a Latin chamber system $\Delta$ has 
universal $2$--cover that is a building if and only if the corresponding loop is a group. 
This is the case, for example, for all the code loops obtained by considering an
isotropic subspace $C\subset V$ of an almost-symplectic $(V,\omega)$ as above.

\smallskip

This implies that a choice of an isotropic subspace $C\subset V$ of an almost-symplectic 
$(V,\omega)$ determines a chamber system that has a building as universal $2$-cover.
We can then formulate the following question. We use the notation $\cB_C$ for the
building obtained in this way.

\medskip

{\bf 6.10.2. Question.} {\it Can the geometric construction of CRSS quantum codes
and of perfect tensors of Theorems~6.6.3.1 and 6.7.4 be used to construct tensor
networks on the buildings $\cB_C$ that satisfy a form of the Ryu--Takayanagi conjecture?}

\medskip

Note that one should not expect in general to have good holographic 
properties for tensor networks on these classes of buildings, and
it is likely that only special cases will satisfy some form of 
Ryu--Takayanagi conjecture, relating
entanglement entropy on the boundary to geodesic lengths in the bulk.
Indeed, it is expected that the $CAT(-1)$ rather than $CAT(0)$ property
may be required for a Ryu--Takayanagi conjecture to hold.
However, even in the absence of these stronger holographic properties, an  
entanglement entropy associated to chamber systems obtained from loop
codes and their perfect tensors would show that there are interesting entangled
states capturing various aspects of the geometry of the chamber system and its 
building universal $2$--cover. 
Properties of tensor networks on
buildings are a topic currently under active investigation in the context of the
holographic AdS/CFT correspondence.

\bigskip

{\bf Acknowledgements.}  N. C. Combe acknowledges support from  the Minerva Fast track grant from 
the Max Planck Institute for Mathematics in the Sciences, in Leipzig.

Yu. Manin acknowledges the continuing strong support from the Max Planck Institute
for Mathematics in Bonn.

M. ~Marcolli  acknowledges support
from NSF grants DMS--1707882 and DMS-2104330.

We thank the referee for suggesting a significant improvement to Section 6.5.

\bigskip 
\centerline{\bf References}

\medskip

[ASh92] M.  Akivis,  A. Shelekhov.  {\it Geometry and algebra of multidimensional
3--webs.} Math. and its Applications 82 (Soviet Series) (1992).

\smallskip

[BoMa07]  D.~Borisov, Yu. Manin. {\it Generalized operads and their inner cohomomorhisms.}
 In: Geometry and Dynamics of Groups
and Spaces (In memory of Aleksander Reznikov). Ed. by M. Kapranov et al.
Progress in Math., vol. 265 (2007), 
Birkh\"auser, Boston, pp. 247--308.
arXiv math.CT/0609748.
\smallskip

[CRSS97] A.R.~Calderbank, E.M.~Rains, P.W.~Shor, N.J.A.~Sloane. {\it Quantum error correction
and orthogonal geometry.} Phys. Rev. Lett. 78 (1997), no. 3 405.

\smallskip

[CRSS98] A.R.~Calderbank, E.M.~Rains, P.W.~Shor, N.J.A.~Sloane. {\it Quantum
error correction via codes over $GF(4)$.} IEEE Transactions of Information Theory, 44 (1998) N.4, 1369--1387.

\smallskip

[Cam03] P. J. Cameron. {\it Chamber systems and buildings}, The Encyclopaedia of Design Theory (2003).

\smallskip

[ChGo90]  O.~Chein, E.~Goodaire. {\it Moufang loops with a unique nonidentity commutator (associator, square).}
J. Alg. 130 (1990), 369--384.

\smallskip

[CoCoNen21]  N.  Combe,  P.  Combe,  H.  Nencka.  {\it Pseudo--elliptic geometry
of a  class of Frobenius manifolds and Maurer--Cartan structures.} 
arXiv:2107.01985.

\smallskip

[CoMa20] N. Combe, Yu. Manin. {\it $F$--manifolds and geometry of information.}
Bull. Lond. Math. Soc. 52 (2020), no. 5, 777--792.  arXiv:math.AG/2004.08808.

\smallskip
[CoMaMar20] N. Combe, Yu. Manin., M. Marcolli. {\it Dessins for modular operad and 
Grothendieck--Teichm\"uler  group.} to appear in ``Topology and Geometry
A Collection of Essays Dedicated to Vladimir G. Turaev", European Mathematical Society, 2021.
arXiv:math.AG/2006.13663.

\smallskip

[CoMaMar21] N.  Combe,  Yu. Manin,  M.  Marcolli. {\it Geometry of information: classical and quantum aspects.} to
appear in Theoretical Computer Science, Special issue for the 70th birthday of Cristian Calude, 2021.

\smallskip

[CoMa21] N. Combe, Yu. Manin. {\it Symmetries of genus zero modular operad.} in ``Integrability, Quantization, 
and Geometry: II. Quantum Theories and Algebraic Geometry", Proceedings of Symposia in Pure Mathematics
Vol.103, pp.101--110, American Mathematical Society, 2021.
arXiv:math.AG/1907.10317.

\smallskip

[Conw85] J.H.~Conway. {\it A simple construction for the Fischer--Griess monster group.} Invent. Math. 79
(1985), 513--540.

\smallskip

[Eil48]  S.~Eilenberg, {\it Extensions of general algebras.} Ann. Soc. Polon. Math., 21 (1948) N.1, 125--134.

\smallskip

[Gri86] R.L.~Griess, {\it Code Loops.} J. of Algebra, 100 (1986) 224--234. 

\smallskip

[GriPir18] A.~Grishkov, R.~Miguel Pires, {\it Variety of loops generated by code loops}, Intern. J. of Algebra and Comp., Vol.~28 (2018) 163--177.

\smallskip

[GuHa09] S.~Gurevich and R.~Hadani, {\it Quantization of symplectic vector spaces over finite fields.}
Journal of Symplectic Geometry 7 (2009), no. 4 475--502.

\smallskip

[GuHa12] S.~Gurevich, R.~Hadani, {\it The Weil representation in characteristic two.} Adv. Math. 
230 (2012), no. 3, 894--926.

\smallskip

[Hall19] J.I.~Hall, {\it Moufang loops and groups with triality are essentially the same thing}. 
Mem. Amer. Math. Soc., Vol.~260 (2019), N.~1252, xiv+186 pp

\smallskip

[HMPS18] M.~Heydeman, M.~Marcolli, S.~Parikh, I.~Saberi,
{\it Nonarchimedean holographic entropy from networks of perfect tensors.} arXiv:1812.04057,
to appear in Advances in Theoretical and Mathematical Physics.

\smallskip

[Hsu00a] T.~Hsu. {\it Moufang loops of class $2$ and cubic forms.}
Math. Proc. Camb. Phil. Soc. 128 (2000), 197--222. 

\smallskip

[Hsu00b] T.~Hsu. {\it Explicit constructions of code loops as centrally twisted products.}
Math. Proc. Camb. Phil. Soc. 128 (2000), 223--232

 \smallskip

[Log93] E.K.~Loginov, {\it On linear representations of Moufang loops.} Commun. 
Algebra, 21 (1993) N.7, 2527--2536.

\smallskip

[Lo69] O. ~Loos. {\it Symmetric spaces. I: General theory.} W. A. Benjamin, Inc. 
New York -- Amsterdam, 1969.

\smallskip

[Mal55]  I. A. Malcev. {\it Analytical loops.} Math. Sbornik 36 (1955), pp. 569--576.

\smallskip

[Ma86] Yu. Manin. {\it Cubic forms.} 2nd Edition, North--Holland, Amsterdam, 1986.

\smallskip

[Ma12] Yu. Manin. {\it A computability challenge: asymptotic bounds
 and isolated error--correcting codes.} WTCS 2012 (Calude Festschrift), ed. by M. . Dinneen
 et al., LNCS 7160 (2012), pp. 174--182. arXiv:1107.4246

 \smallskip

[MaMar12] Yu. Manin, M. Marcolli. {\it Kolmogorov complexity and the asymptotic bound 
for error--correcting codes.} Journ. of Diff. Geometry 97 (2014), pp. 91--108.
arXiv:1203.0653v2

\smallskip

[MaMar19] Yu. Manin, M. Marcolli. {\it Nori diagrams and persistent homology.}
Math. Comput. Sci. 14 (2020), no. 1, 77--102.

\smallskip

[Mar19]  M. Marcolli.   {\it Gamma spaces and information.} Journ. of Geometry and Physics,
140 (2019), pp. 26--55.

\smallskip

[MeiStWe13] U.~Meierfrankenfeld, G.~Stroth, R.M.~Weiss. {\it Local identification 
of spherical buildings and finite simple groups of Lie type.} Math. Proc. Camb. Phil. Soc., Vol.154 (2013),
pp.  527--547.

\smallskip

[MoChe91]  E.  Morozova,  N.  Chentsov.  {\it Markov invariant geometry on state
manifolds.} J.  Soviet Math,  56 (1991),  pp. 2648--2669. (Russian original 1989).

\smallskip

[Na88] P.  T.  Nagy. {\it Invariant tensorfields and the canonical connection of a 3--web.}
Aequationes Math., 35 (1988), pp. 31--44.

\smallskip

[Na92] P. T. Nagy. {\it Moufang loops and Malcev algebras.} Seminar Sophus Lie, 3 (1992),
pp. 65--68.

\smallskip

[Na08] G. P.~Nagy. {\it Direct construction of code loops.}
Discrete Mathematics, 308 (2008) 5349--5357.

\smallskip

[NaRob21] B.~Nagy, D.M.~Roberts. {\it (Re)constructing Code Loops.}
The American Mathematical Monthly, 128 (2021) N.2, 151--161.

\smallskip

[OtiSta17] A.~Otiman, M.~Stanciu, {\it 
Darboux--Weinstein theorem for locally conformally symplectic manifolds}. 
J. Geom. Phys. 111 (2017), 1--5.

\smallskip

[Pa03] E. Paal. {\it Moufang loops and Lie algebras.} 
Czechoslovak J. Phys. 53:11 (2003),  pp.  1101--1104
arXiv:math--ph/0307014 

\smallskip

[PYHP15] F.~Pastawski, B.~Yoshida, D.~Harlow, J.~Preskill. {\it Holographic quantum error-correcting codes: Toy models for the bulk/boundary correspondence.} JHEP 06 (2015),  pp.  149 -- 204.

\smallskip
[PeSuWeiZa20] M.~Petrera, Yu. B. Suris, Kangning Wei, R. Zander.
{\it Manin involutions for elliptic pencils and discrete integrable systems.}
 Math. Phys. Anal. Geom. 24 (2021), no. 1, Paper No. 6, 26 pp. 
arXiv:2008.08308.
 
\smallskip

[Sa61] A.  A.  Sagle. {\it Malcev algebras.} Trans. AMS, 101 (1961), pp. 426--458.

\smallskip

[Shult11] E.E.~Shult. {\it Points and Lines. Characterising the Classical Geometries}, Springer (2011).

\smallskip

[SpVe00]  T.  A.  Springer,  F.  D.  Veldkamp. {\it Octonions, Jordan algebras, and 
exceptional groups.}  Springer Verlag, Berlin (2000).

\smallskip

[Vi63] E. Vinberg. {\it The theory of homogeneous convex cones.} Trans. Moscow
Math. Soc. 12 (1963), pp. 340--403.

\enddocument